\documentclass[
 aip,
 amsmath,amssymb,
 reprint,%
numerical,%
]{revtex4-2}

\usepackage{graphicx}
\usepackage{dcolumn}
\usepackage{bm}
\usepackage{subcaption}
\usepackage{caption}
\usepackage{titlesec}
\renewcommand{\thesection}{\Roman{section}}
\usepackage[utf8]{inputenc}
\usepackage[T1]{fontenc}
\usepackage{mathptmx}
\usepackage{etoolbox}
\usepackage{subcaption}
\usepackage{siunitx}
\usepackage{pgfplots}
\usepackage{hyperref}
\usepackage{tikz}
\usepackage{pdfpages}
\pgfplotsset{compat=1.18}
\usepackage[justification=raggedright,singlelinecheck=false]{caption}
\usepackage{ragged2e}
\usepackage[justification=justified,singlelinecheck=false]{caption}
\titleformat{\section}
  {\centering\fontsize{14pt}{10pt}\bfseries}
  {\thesection.}{0.3em}{} 
\titlespacing*{\section}{0pt}{12pt}{6pt}  

\usepackage{natbib}
\setcitestyle{numbers,square}
\usepackage{hyperref}
\hypersetup{
    colorlinks=true,
    linkcolor=blue,
    citecolor=blue,
    urlcolor=blue
}

\makeatother

\makeatletter
\AtBeginDocument{\let\LS@rot\@undefined}
\makeatother

\begin{document}

\title{\huge Optimizing defect states in $(Bi_{0.3}Sb_{0.7})_{2}Te_{3}$ ternary topological insulators using indium doping}  

\author{Kanav Sharma}
\affiliation{Department of Physical Sciences, Indian Institute of Science Education and Research Kolkata, Nadia, 741246, West Bengal, India}

\author{Ritam Banerjee}
\affiliation{Department of Physical Sciences, Indian Institute of Science Education and Research Kolkata, Nadia, 741246, West Bengal, India}

\author{Anuvab Nandi}
\affiliation{Department of Physical Sciences, Indian Institute of Science Education and Research Kolkata, Nadia, 741246, West Bengal, India}

\author{Radha Krishna Gopal}
\affiliation{Department of Physics and Material Sciences and Engineering, Jaypee Institute of Information Technology, Sector 62, Noida, India}

\author{Chiranjib Mitra}
\affiliation{Department of Physical Sciences, Indian Institute of Science Education and Research Kolkata, Nadia, 741246, West Bengal, India}
\email{chiranjib@iiserkol.ac.in}

\date{\today}

\begin{abstract}
\large
\raggedright This study investigates the influence of indium doping on the defect states in $(Bi_{0.3}Sb_{0.7})_{2}Te_{3}$ (BST) ternary topological insulators. Thin (10 nm) and thick (60 nm) films of pristine BST and indium-doped BST ($In_{0.14}(Bi_{0.3}Sb_{0.7})_{1.86}Te_{3}$) were synthesized using pulsed laser deposition. The electronic properties were characterized through low-frequency noise spectroscopy and temperature-dependent resistance (R-T) measurements. For the 10 nm films, R-T analysis revealed that indium doping shifts the thermal activation energy by approximately 100 meV. This doping also suppresses a shallow impurity band at 72 meV, a finding corroborated by 1/f noise measurements. In the 60 nm films, noise spectroscopy was used to probe deep defect states, where indium doping was found to increase the activation energy from 292.3 meV to 392 meV—a consistent shift of 100 meV. These findings demonstrate that indium doping is an effective method for systematically modifying both shallow and deep defect states, enhancing the insulating properties and offering a mechanism to engineer the electronic behavior of topological insulators for advanced electronic applications where noise reduction is crucial. 
\end{abstract}

\maketitle

\section{INTRODUCTION}
\large
Topological insulators\cite{colloquium,Kane2005,t3,t4,t5,t5,t6,t7,t8} are a distinct class of novel phases of quantum matter that host helical surface states in the inverted bulk band gap. These surface states exhibits linear dispersion and owing to their helical nature they form Dirac cones. Depending on the number of surface Dirac cones, these topological states are further classified into weak and strong topological insulators. These phases of topological insulators are distinguished by distinct quantum numbers and the one we have studied here are classified as $Z_2$ invariant topological insulators. Though surface states in weak topological insulators can be localised by strong disorder, the topologically protected Dirac states in strong TIs cannot be localised by non-magnetic disorder\cite{strong_weak_TI}. This unique topological protection and immunity against backscattering from disorder makes this class of materials distinct from other 2D materials such as graphene and transition metal dichalcogenides\cite{TR,TR1}. The robustness of the surface states makes these classes of materials highly promising for applications in quantum computing\cite{superconducting}, spintronics\cite{spintronics,pal2024enhancement}, and low-noise electronics\cite{low_noise}. 

Over the last decade or so, the research on the topological quantum materials has gained enormous interest worldwide owing to their exotic transport and optical properties. These exotic phenomena include the Dirac fermion-dominant half-integer (per surface) quantum Hall effect in $BiSbTeSe_{2}$(BSTS) exfoliated flakes, the quantised anomalous Hall effect in magnetic and antiferromagnetic topological insulators, superconducting proximity-driven Majorana modes, and topological axionic states with zero (integer) Chern number. However, these studies are mainly focused on the devices with an ultraflat surface with minimal surface or bulk disorder. Very few TI materials exists that show highly bulk-insulating and surface-dominant transport simultaneously\cite{kushwaha2016sn,PhysRevMaterials.3.054204}. These devices are either grown by the molecular beam epitaxy (MBE)—known for growing very high-quality and minimally disordered TI thin films—or by exfoliating flakes from the high-quality single crystals such as BSTS and Sn-doped BSTS\cite{flakes}. Therefore, these TI devices, which are very clean in nature, are not the ideal platform for testing of possible realisation of Z2 topological protection of surface Dirac fermions against strong disorder in a TI.

On the other hand, pulsed laser deposition (PLD)\cite{pulsed,pld2,mitra2002magnetotransport} technique naturally provides a unique platform to test this topological protection of surface Dirac fermions. These thin films, being granular and polycrystalline in nature, are heavily disordered in comparison to MBE grown thin films and exfoliated flakes. In PLD-grown films, in addition to the vacancies and antisite disorder, the randomly orientated grain boundaries strongly restrict the conduction process at the surface and in the bulk. It has been observed experimentally in low-temperature transport experiments that even 2D thin films of metals show metal-insulator transition due to the granular nature owing to the quantum confinement in the potential barrier between nano-sized grains\cite{granular}. However, due to the intrinsic topological protection, surface states are expected to overcome the grain boundary potential and other impurities (Te, Bi, and Sb antisite defects) and remain extended or delocalised in nature. It could be a possible physical reason behind the high thermoelectric efficiency in the Bi- and Sb-based thermoelectric materials\cite{thermoelectric}. The presence of such impurities or grain boundaries in PLD grown thin films sets our study apart from many other such studies carried out on the high-quality thin films. Other two-dimensional systems like graphene and van der Waals (vdW) materials exhibit weak or strong localisation depending on the nature of the disorder, be it intrinsic (grain boundaries or other point defects) or extrinsic (substrate-dependent) disorder.

Low frequency noise spectroscopy, viz., 1/f noise\cite{hooge,n2,n3,n4} is a powerful experimental tool that can probe different types of scattering phenomena or dephasing mechanisms caused by various forms of intrinsic or extrinsic disorder in electronic devices\cite{islam2023benchmarking,pal2011microscopic,balandin2013low,rehman2023low}. This technique helps in identifying and resolving the challenges exhibited due to the presence of various kinds of disorders in order to improve device performance. In TIs, the dominant source of 1/f noise has been identified to be multiple impurity bands formed by antisite defects\cite{antisite,antisite2,antisite3} (Bi/Te or Sb/Te) and vacancies (Se or Te), forming shallow impurity bands in the bulk band gap and charged puddles near the chemical potential\cite{skinner2012bulk}. In topological insulator (TI) films, Coulomb disorder--induced charge puddles in the bulk have been shown to affect surface transport, even in bulk-insulating alloys such as BiSbTeSe$_2$\cite{PhysRevB.96.195135}. While previous research has primarily focused on metallic $(Bi,Sb)_{2}Te_{3}$ thin films\cite{islam2017bulk}, this work extends noise spectroscopy analysis to $(Bi_{0.3}Sb_{0.7})_{2}Te_{3}$ (BST) and indium-doped $In_{0.14}(Bi_{0.3}Sb_{0.7})_{1.86}Te_{3}$ (IBST) samples at temperatures as low as 90K (liquid $N_2$ temperature). We investigated two sets of films each for BST and IBST. The first set, with a thickness of 10nm, is classified as thin in this study, while the second set, having a thickness of 60nm, is referred to as thick films. Comparative noise spectra of the thinner films of BST and IBST reveal that indium doping effectively suppresses the impurity band. The R-T data of this set also shows a notable shift of activation energy  by 100 meV in the doped sample (IBST). Similarly, in thick films, indium doping induces an equivalent 100 meV shift in the activation energy of defect states; which however, was derived from the noise spectra analysis instead of the R-T data. It is important to note that for thin films the bulk contribution is significantly reduced in comparison to thick films. Therefore the effect of disorder contribution is more perceptible in thicker films. These findings emphasize the consistent impact of indium doping in both the thin and thick films, offering valuable insights into its role in modifying electronic properties.

\section{EXPERIMENTAL DETAILS}
Thin films of $(Bi_{0.3}Sb_{0.7})_{2}Te_{3}$  and indium-doped $In_{0.14}(Bi_{0.3}Sb_{0.7})_{1.86}Te_{3}$ were grown on silicon substrates using the pulsed laser deposition (PLD, KrF-excimer laser, wavelength-248nm) technique by ablating individual targets of stochiometric $(Bi_{0.3}Sb_{0.7})_{2}Te_{3}$(BST) and $In_{0.14}(Bi_{0.3}Sb_{0.7})_{1.86}Te_{3}$(IBST)\cite{pulsed}. All the other details of the conditions of growth of these thin films have been given in the supplementary material. The resistance fluctuation measurements were carried out in the low-bias current range where $S_V$(f)(Power spectral density) was linear in V(voltage across the sample). The supplementary section contains comprehensive details regarding the experimental setup, noise measurement procedure, data acquisition process and sample characterisation.

\begin{figure}[!htbp]
    \centering
    
    \begin{subfigure}{0.5\textwidth}
        \centering
        \includegraphics[width=\linewidth]{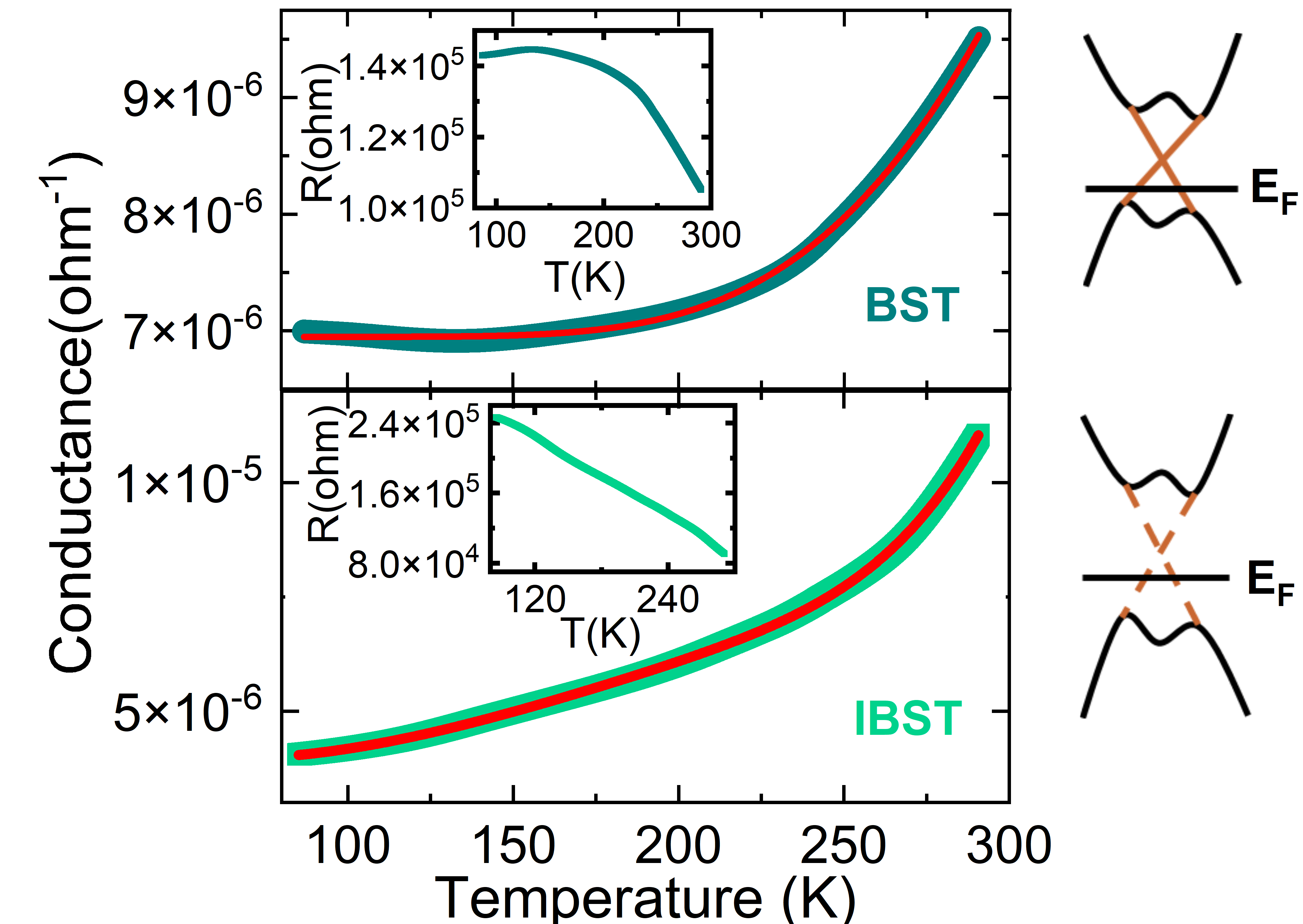}
        \vspace{-15pt}
        \captionsetup{font=Large}
        \captionsetup{justification=centering}
        
        \caption{}
        \label{fig:1a}
    \end{subfigure}
    \vspace{0.1cm}
    \hfill
    
    \begin{subfigure}{0.5\textwidth}
        \centering
        \includegraphics[width=\linewidth]{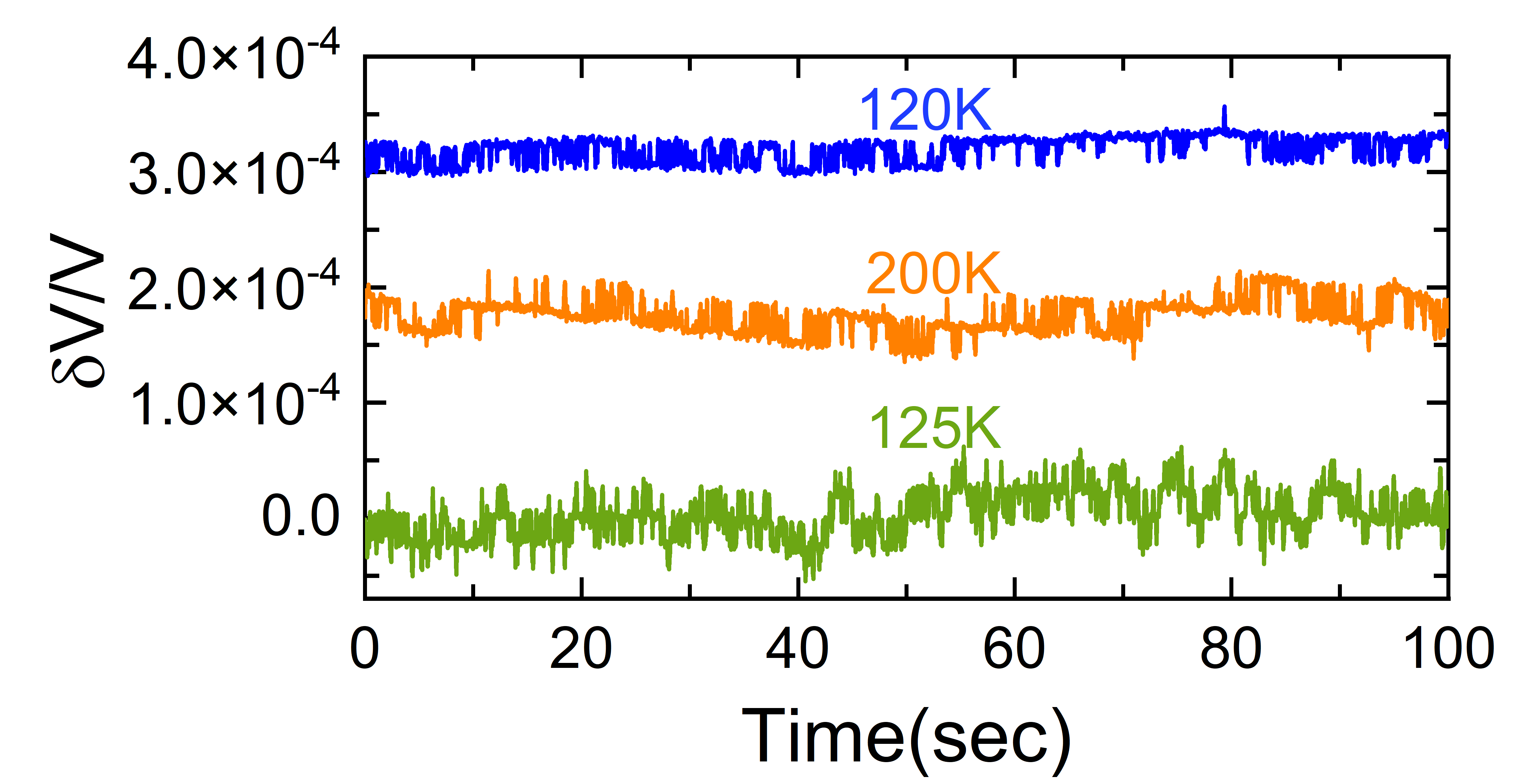}
        \vspace{-18pt}
        \captionsetup{font=Large}
        \captionsetup{justification=centering}
        \caption{}
        \label{fig:1b}
    \end{subfigure}
    \vspace{0.1cm}
    \hfill
    
    \begin{subfigure}{0.5\textwidth}
        \centering
        \includegraphics[width=\linewidth]{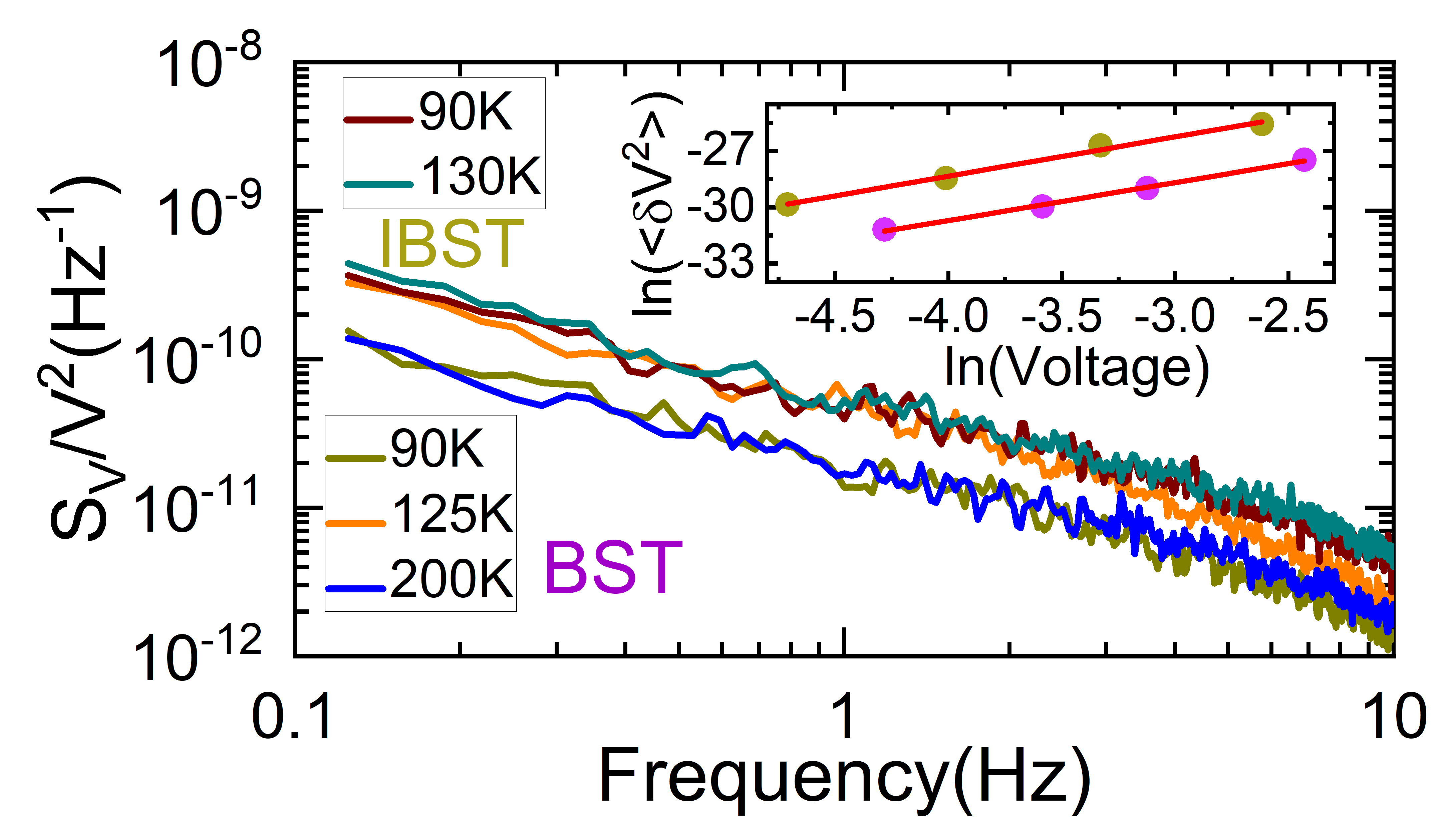}
        \vspace{-20pt}
        \captionsetup{font=Large}
        \captionsetup{justification=centering}
        \caption{}
        \label{fig:1c}
    \end{subfigure}
    
    \vspace{15pt}
    \caption{\justifying Figures (a), (b), and (c) represent the following for thin samples: 
    (a) The conductance versus temperature graphs for BST and IBST thin films are presented, with insets displaying the corresponding resistance versus temperature plots. Adjacent to these, band diagrams are shown on the right. Within the band gap, solid orange lines represent surface states, whereas, the dashed orange line represents a decrease in surface state contribution. (b) Normalized voltage fluctuations for BST at 125 K (green) and 200 K (orange), and for IBST at 120 K (blue), as a function of time, shifted along the vertical axis for clarity.
    (c) Normalised power spectral density versus frequency at various temperatures, with inset showing log-log plots of integrated variance versus voltage across the sample for both BST and IBST.}
\end{figure}

\section{RESULTS AND DISCUSSION}
Fig.~\ref{fig:1a} shows conductance versus temperature graphs for BST and IBST thin films with insets displaying the corresponding resistance versus temperature plots. Adjacent to these, band diagrams are shown on the right panel. Within the band gap, solid orange lines represent surface states, whereas, the dashed orange line depicts a decrease in surface state contribution due to doping. The temperature dependence of resistance (R-T) of the thin films exhibits highly bulk insulating character, as is shown in inset of Fig.~\ref{fig:1a}. The value of the resistance for the two thin films, BST and IBST, reaches 145 kOhms and 240 kOhms at the lowest achievable temperature of 90 K. There is a clear indication from the R-T data that the two thin films exhibit distinct characteristics. The R-T data of BST film reveals a bulk insulator-to-metal transition at a temperature as high as 130 K, which can be seen by the downturn in the R-T data around this temperature. This indicates a dominant surface transport with the Fermi level positioned near the Dirac point within the bulk band gap. On the contrary, in the case of IBST, one can see a monotonically increasing trend in the R-T data, with no downturn in the low-temperature regime, which in turn can be attributed to the enhanced localisation of the bulk electrons as well as decreased contribution from the surface states due to reduced spin-orbit coupling\cite{brahlek1}. It has been observed in the ARPES, time-domain terahertz, and transport experiments that substitution of indium changes the topology and spin helical texture of the surface states, thus making a topological quantum phase transition at certain indium concentrations of around 6-7\% \cite{brahlek1,brahlek2,transport_In}. In spite of the fact that these BST thin films are more disordered than the MBE-grown thin films and single crystal flakes, as seen from the surface morphology in the SEM data shown in the supplementary section, it is remarkable to observe the dominant surface transport at a temperature as high as 130K. 

In order to estimate the contribution of the surface and bulk states in the R-T data we have fitted the conductance (G) data as a function of temperature to the parallel resistor model(in other words series conduction model)\cite{Syer_fit,PbSnTe_fit,mitra2001growth}. This model considers bulk and surface conducting channels as two separate parallel conducting channels in moderately bulk-insulating topological insulators. The total conductance in this model can be written as follows:
\begin{equation} 
\text{\large{$G_{\text{tot}} = G_{\text{sur}} + G_{\text{bulk}}$}} \tag{1} 
\end{equation} 
$G_{\text{sur}} = \frac{1}{a}$ represents the surface channel conductance. For thin BST $G_{\text{bulk}} = \frac{1}{c* \exp\left(\frac{\Delta E}{k_bT}\right)}$ represents the bulk channel conductance. For thin IBST $G_{\text{bulk}} = \frac{1}{c* \exp\left(\frac{\Delta E}{k_bT}\right)} + \frac{1}{d* \exp\left(\frac{\Delta E_1}{k_bT}\right)}$. Here, a, b, c, d, $ \Delta E$, and $ \Delta E_1$ are the fitting parameters. The data fitted with this model for the two thin films are shown in figure 1(a). The fitting results reveal that the activation energy $\Delta E$ for BST is 143 meV, while for IBST, $\Delta E$ and $\Delta E_1$ are 36 meV and 264 meV, respectively. A table (Table I) has been included to comprehensively display the activation energies and surface contributions for all samples. The significantly high value of $\Delta E_1$ could be attributed to either an increase in the bulk band gap or to the presence of a deep impurity band. Whereas, $\Delta E$ could be the effect of charged impurities causing potential fluctuations leading to meandering of band edges\cite{skinner2012bulk}. Also in Lostak et al. 1993\cite{suppression}, the authors have found that the nature of carriers taking part in the conduction are holes; thus, indium doping results in the reduction of the number of holes, causing the sample to become more insulating. This is consistent with the enhanced $\Delta E_1$ mentioned above. The R-T fitting results indicate that surface conduction accounts for 73\% in BST and 36\% in IBST. Indium doping plays a pivotal role: as a lighter element, it may suppress spin-orbit coupling within the system, thereby diminishing the topological protection of surface states. This idea is captured in the band diagram shown in Fig.~\ref{fig:1a} , where the dashed (orange line) represents the surface states with reduced topological protection. This picture captures the reduced contribution of surface states as seen from the parallel resistor model.

\begin{figure}[!htbp]
    \centering
    
    \begin{subfigure}{0.48\textwidth}
        \centering
        \includegraphics[width=\linewidth]{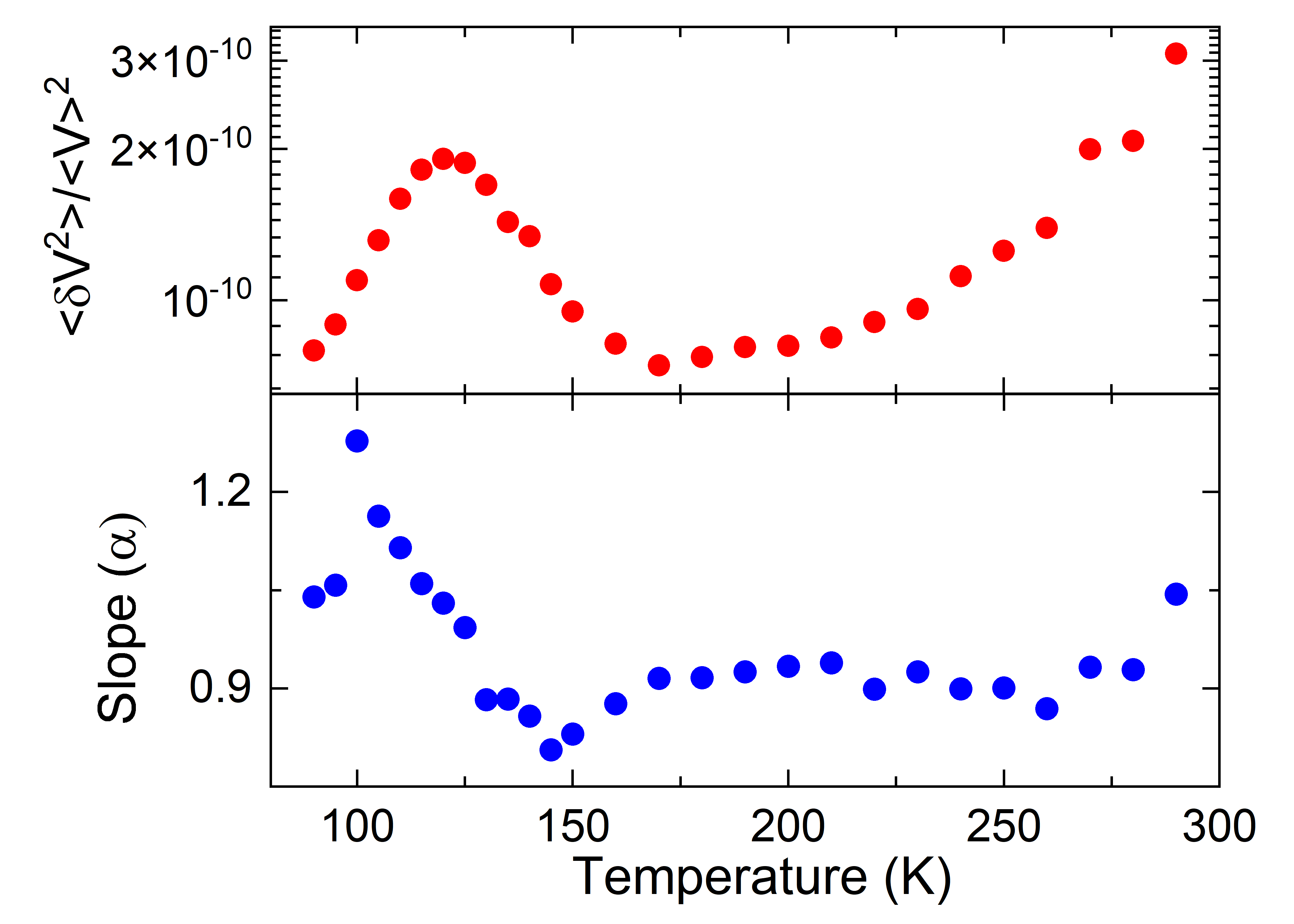}
        \vspace{-16pt}
        \captionsetup{font=Large} 
        \captionsetup{justification=centering}
        \caption{}
        \label{fig:2a}
    \end{subfigure}
    \vspace{0.6cm}
    \hfill
    
    \begin{subfigure}{0.48\textwidth}
        \centering
        \includegraphics[width=\linewidth]{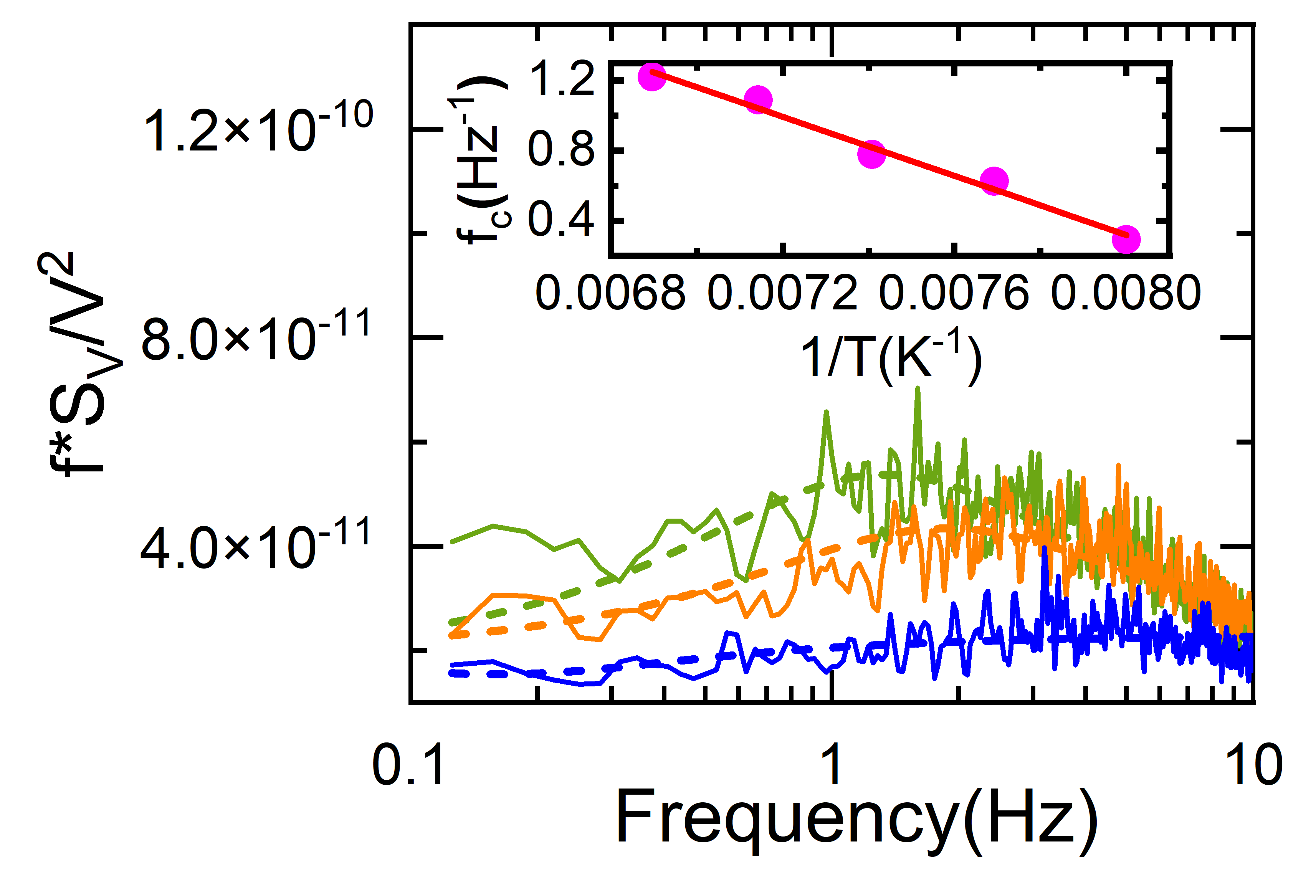}
        \vspace{-22pt}
        \captionsetup{font=Large} 
        \captionsetup{justification=centering}
        \caption{}
        \label{fig:2b}
    \end{subfigure}
    \vspace{0.6cm}
    \hfill
    
    \begin{subfigure}{0.48\textwidth}
        \centering
        \includegraphics[width=\linewidth]{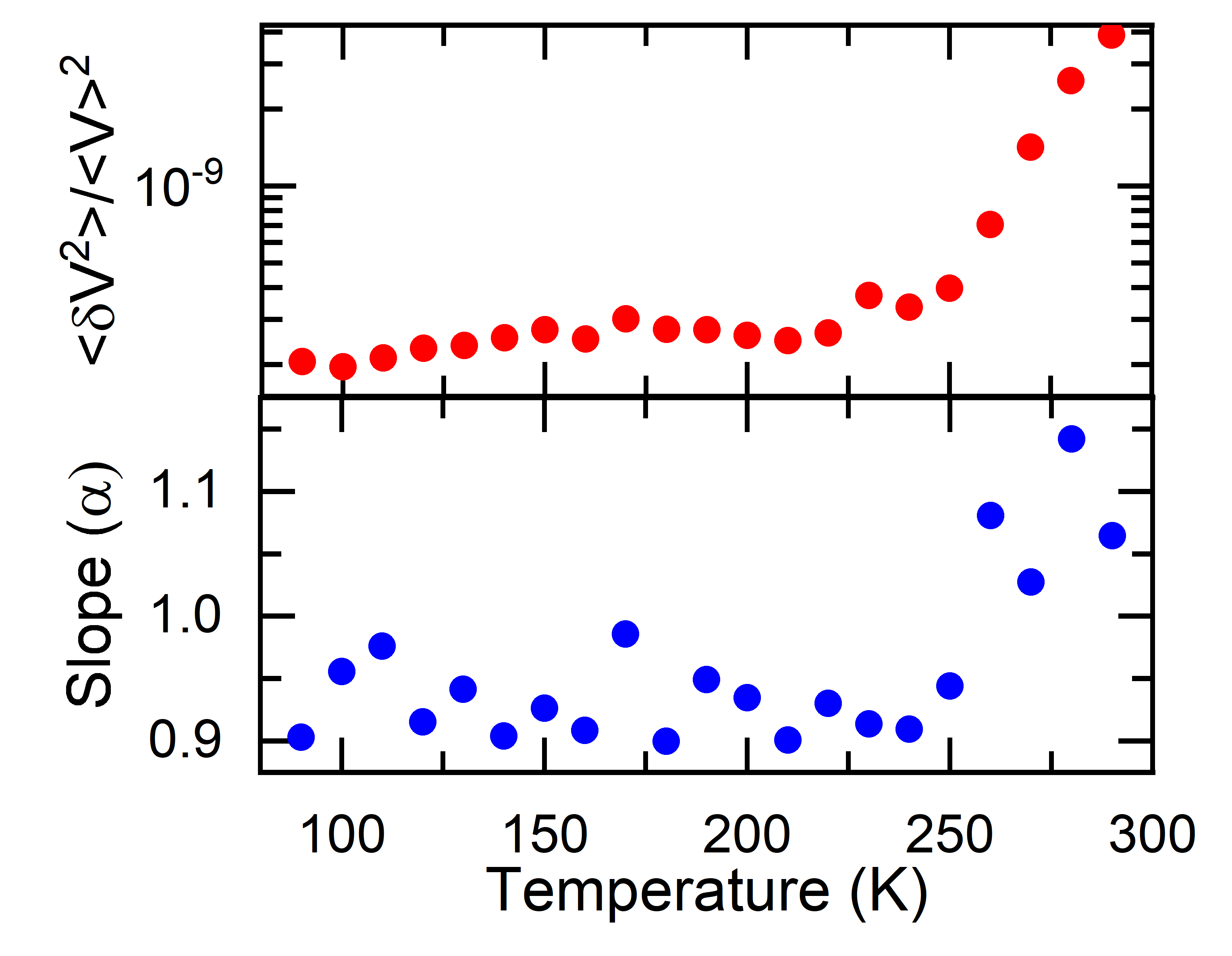}
        \vspace{-22pt}
        \captionsetup{font=Large} 
        \captionsetup{justification=centering}
        \caption{}
        \label{fig:2c}
    \end{subfigure}
    
    \vspace{15pt}
    \caption{\justifying Figures (a), (b), and (c) represent the following for thin samples: (a) Normalized relative variance and corresponding slope as a function of temperature for the BST sample. (b) Lorentzian~+~$1/f$ fits for the BST sample at 125~K (green) and 135~K (orange), each yielding well-defined $f_c$ values, and at 200~K (blue) where the fit is dominated by the $1/f$ component (parameter $A$ in Eq.~4), preventing a meaningful determination of $f_c$. Inset: Arrhenius fit of $f_c$ from the valid temperature range. (c) Normalized relative variance and corresponding slope as a function of temperature for the IBST sample.
}
\end{figure}

\begin{table*}[t] 
\centering
\renewcommand{\arraystretch}{1.5} 
\begin{tabular}{|>{\centering\arraybackslash}m{3cm}|>{\centering\arraybackslash}m{3cm}|>{\centering\arraybackslash}m{3cm}|>{\centering\arraybackslash}m{3cm}|>{\centering\arraybackslash}m{3cm}|}
\hline
Sample(with thickness in nm) & Activation energy(meV)(from R-T data) & Activation energy(meV)(from 1/f noise data) & Surface contribution(\%) from R-T data \\ \hline
BST(10) & 143 & 72 & 69  \\ \hline
IBST(10) & 264 & - & 32 \\ \hline
BST(60) & 58 & 292.3 & 69.4  \\ \hline
IBST(60) & 61.2 & 392 & 71  \\ \hline
\end{tabular}
\vspace{5pt}
\caption{\justifying The table displays the activation energies of various samples determined using R-T and 1/f noise, along with the percentage of surface state contribution for each sample.}
\end{table*}
Fig.~\ref{fig:1b} shows the normalised voltage time series data at three different temperatures. In Fig.~\ref{fig:1c}, the normalised power spectral density (PSD($S_V$)) with frequency is depicted for the thin films measured at different temperatures. The log-log plot of integrated PSD versus voltage across the sample shown in the inset of Fig.~\ref{fig:1c}, exhibits a slope of 2 for both the thin films, indicating that both the samples are in ohmic regime\cite{ohm_law}. The magnitude of the noise is obtained by calculating the integrated relative variance given by\cite{biswas2019resistance},
\begin{equation} 
\text{\large{$\frac{\langle \delta V^2 \rangle}{\langle V \rangle^2} = \frac{1}{\langle V \rangle^2} \int_{0.1}^{10} S_V(f) \, df$}} \tag{2} 
\end{equation} 
Fig.~\ref{fig:2a} upper panel shows the temperature-dependent relative variance, where we observe the signature of generation-recombination noise around 110K. The slope of the PSD versus frequency, represented by $\alpha$, ranges from 0.8 to 1.3 throughout the entire temperature range, as illustrated in the lower panel of Fig.~\ref{fig:2a}. Notably, the increase in slope is characteristic of generation-recombination noise. The generation and recombination is due to excitation of carriers between two states i. defect state and valence band. To determine the slope $\alpha$, we employed an equation of the form:
\begin{equation} 
\text{\large{$S_V = \frac{A}{f^{\alpha}} ,$}} \tag{3} \end{equation} 
where A is the fitting parameter. In the region of generation-recombination, the Lorentzian component is added to the 1/f part given by:
\begin{equation} 
\text{\large{$f \cdot S_V = A + \frac{B \cdot f_c}{f^2 + f_c^2} \quad$}} \tag{4} 
\end{equation} 
where A and B are the fitting parameters. We successfully determined the corner frequencies ($f_c$), which represent the switching frequency between the two states. In Fig.~\ref{fig:2b}, we display two sets of data fits at 125K (green) and 135K (orange), each with its respective fit. For comparison, the 200~K data (blue) are shown with the same Lorentzian~+~$1/f$ fit; although the fit matches the spectral shape, it is dominated by the $1/f$ component (parameter $A$ in Eq.~4), preventing a meaningful determination of $f_c$ in this regime. Subsequent analysis involved tracking the shift in $f_c$ with increasing temperature\cite{freq_shift}. The extracted corner frequencies are plotted as a function of inverse temperature in the inset of Fig.~\ref{fig:2b} and fitted to the Arrhenius equation of $f_c$ given by,
\begin{equation} 
\text{\large{$f_c = f_o \exp\left(-\frac{\Delta E}{k_B T}\right)$}} \tag{5} 
\end{equation}

\begin{figure}[t]
    \centering
    
    \begin{subfigure}{0.45\textwidth}
        \centering
        \includegraphics[width=\linewidth]{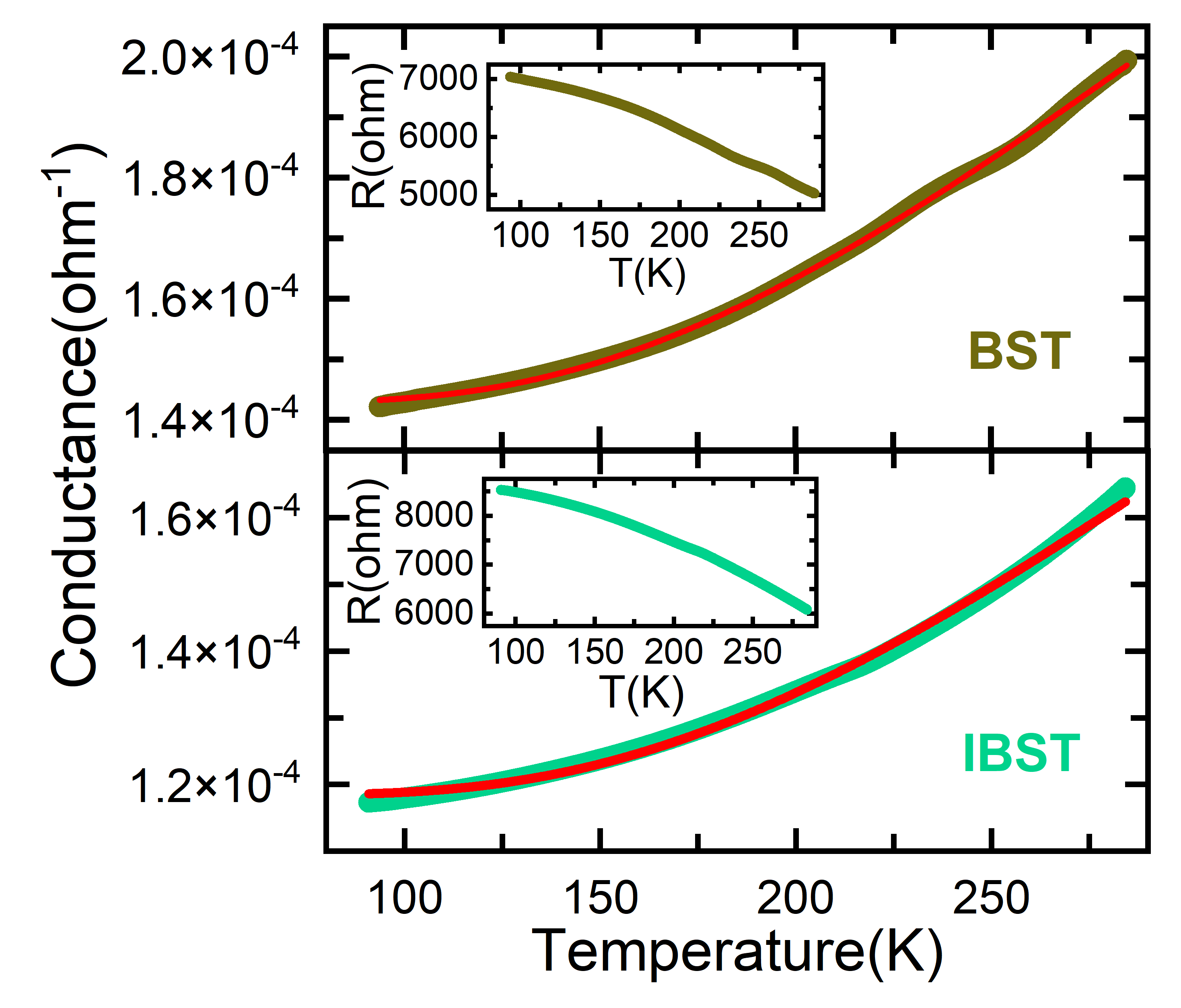}
        \vspace{-22pt}
        \captionsetup{font=Large} 
        \captionsetup{justification=centering}
        \caption{}
        \label{fig:3a}
    \end{subfigure}
    \vspace{0.6cm}
    \hfill
    
    \begin{subfigure}{0.45\textwidth}
        \centering
        \includegraphics[width=\linewidth]{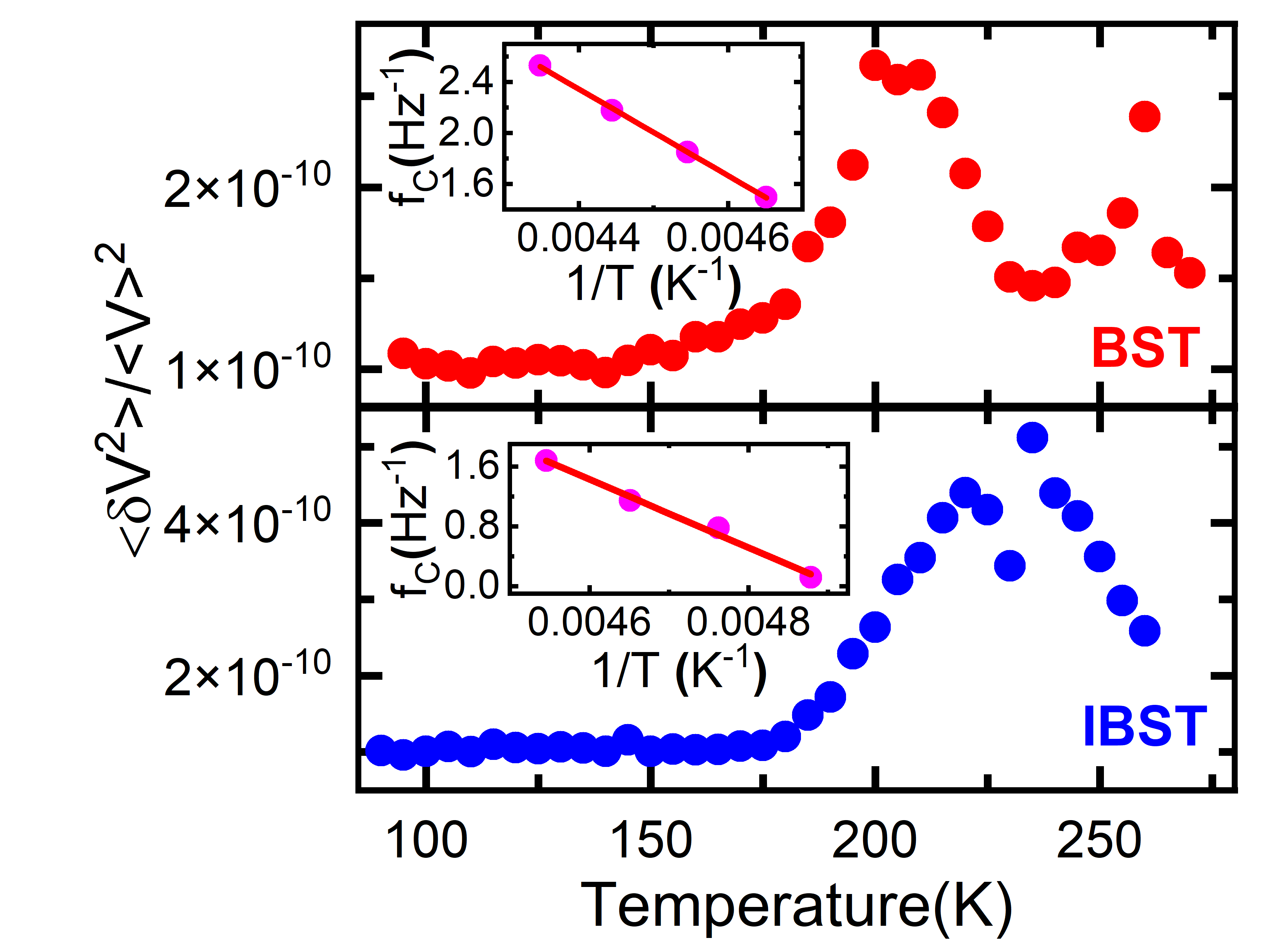}
        \vspace{-20pt}
        \captionsetup{font=Large} 
        \captionsetup{justification=centering}
        \caption{}
        \label{fig:3b}
    \end{subfigure}
    
    \vspace{15pt}
    \caption{\justifying Figures (a) and (b) illustrate the following for thick samples: (a) Conductance versus temperature for BST and IBST, with insets showing resistance versus temperature for each. (b) Normalized relative variance as a function of temperature for BST and IBST, with insets for each showing the fitting to the Arrhenius $f_c$ equation.}
\end{figure}

By fitting the data to equation 5, the impurity band energy of the thin BST film was found to be 72 meV, indicating that impurity band excitations are dominant. This arises from a shallow impurity band in the BST thin film, induced by the presence of vacancies\cite{biswas2019resistance,110K_peak}.  To reduce these defects, we doped BST with indium, which causes more ionic polar bonds between In-Te and hence reduces the number of defects states.\cite{linear_antisite,antisite_vacancies,suppression}. Fig.~\ref{fig:2c} upper panel shows a complete absence of generation-recombination noise, keeping the baseline magnitude of noise the same as that of BST. The slope $\alpha$ for the IBST sample is also shown in Fig.~\ref{fig:2c} lower panel and varies between 0.9 and 1.15. In Fig.~\ref{fig:2a} upper panel, there is a monotonic increase in magnitude of noise from 200 K onward for BST. This effect is attributed to a strengthened coupling between charge carriers and phonons\cite{phonons,phonon2}. This enhanced coupling arises due to the heightened anharmonicity present in the sample in this temperature regime. However, in IBST this monotonic increase sets at 250 K, as shown in Fig.~\ref{fig:2c} upper panel, which might be an indication of a reduction of this coupling. This might be on one hand due to  indium doping, on the other it could  due to decrease in bismuth content which is primarily responsible for the anharmonicity.

\begin{figure}[t]
    \centering
    \includegraphics[width=0.45\textwidth]{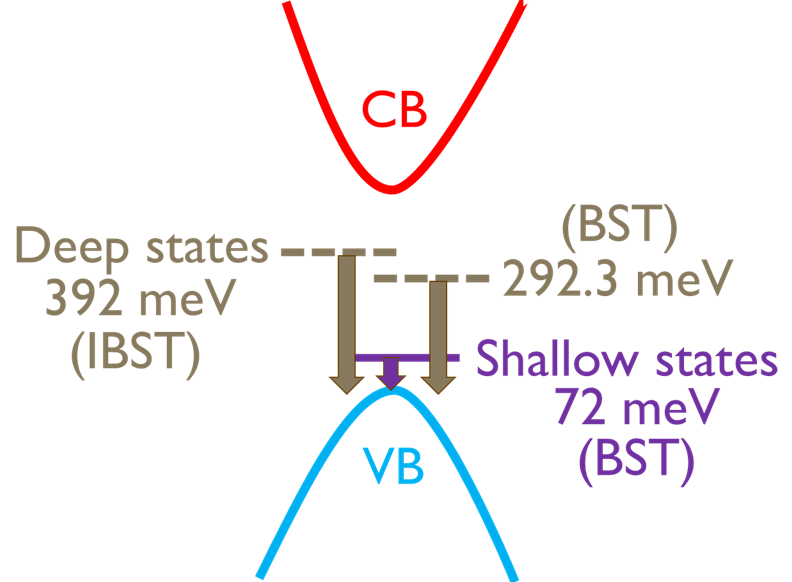}
    \caption{\justifying  Schematic band diagram showing the position of shallow and deep defect states in the bulk band gap, as obtained from low-frequency noise measurements for both BST and IBST films. Shallow defects (blue) near the valence band lead to lower activation energies, while deep defects (brown) originating from antisite disorder, exhibit higher activation energies. The relative positions are drawn for illustration and are not to scale.}
    \label{fig:4}
\end{figure}

For the thicker BST and IBST samples, the R–T data shown in Fig.~\ref{fig:3a} were fitted with the parallel resistor model, yielding activation energies of 58 meV and 61.2 meV, respectively. Increasing sample thickness or doping with indium leads to higher impurity density, promoting the formation of charge puddles. These activation energies have been theoretically attributed to the formation of electron-hole puddles\cite{skinner2012bulk,shklovskii2013electronic} and has also been established experimentally by our group and is part of a separate study\cite{sharma2025bandmeanderingchargedimpurity}. At elevated temperatures, the two generation-recombination noises appear nearly identical as shown in Fig.~\ref{fig:3b}, hence we focused on analysing the generation-recombination noise near 210 K for both samples. The activation energies of defect states, were obtained from the fits of equation (5) to the data shown in the insets of Fig.~\ref{fig:3b}. The values were found to be 292.3 meV for BST, increasing to 392 meV for IBST—an enhancement of 100 meV, consistent with the R–T results for thin films. These energies are attributed to antisite defects, whose formation in Bi– and Sb–tellurides is favored by low bond polarity and can be modified by dopant-induced changes in bond polarity\cite{linear_antisite}. In particular, indium incorporation increases bond polarity, thereby raising the antisite defect formation energy. Using the band diagram in Fig.~\ref{fig:4}, we schematically illustrate the location of the deep and shallow defect states within the bulk band gap, as extracted from our low-frequency noise spectroscopy analysis. The shallow defects, such as vacancy-induced impurity bands, are located closer to the valence band edge and are responsible for generation–recombination features at lower activation energies (e.g., 72 meV in thin BST film). The deep defects, associated with antisite disorder, lie deeper in the gap (e.g., ~292–392 meV) and dominate in thick films. These defect energies and their shifts upon indium doping highlight how the defect landscape can be engineered to tune electronic transport.

\section{CONCLUSION}
In summary, our study highlights the significant influence of shallow and deep defect states on the behavior of topological insulators. We can either compensate for these defects or enhance the activation energy of defect states through targeted doping, an effect observed in both the thin and thick films. Most importantly we observed a 100 meV shift in the activation energy as obtained from R-T measurements of thin films, along with a comparable change in impurity band excitation derived from the noise spectroscopy of thick films. This highlights the role of indium in ternary BST samples. These findings further demonstrate indium's effectiveness in enhancing the electronic properties of topological insulators, making them well-suited for advanced electronic applications where noise reduction is crucial.\\

\begin{center}
{\Large\bfseries ACKNOWLEDGEMENTS}
\end{center}
The authors acknowledge the constant support of Professor Bhavtosh Bansal and Associate Professor Atindra Nath Pal. The authors acknowledge the University Grant Commission (UGC) of the Government of India for financial support.

\bibliographystyle{aipnum4-2}
\bibliography{aipsamp}

\begin{thebibliography}{54}%
\makeatletter
\providecommand \@ifxundefined [1]{%
 \@ifx{#1\undefined}
}%
\providecommand \@ifnum [1]{%
 \ifnum #1\expandafter \@firstoftwo
 \else \expandafter \@secondoftwo
 \fi
}%
\providecommand \@ifx [1]{%
 \ifx #1\expandafter \@firstoftwo
 \else \expandafter \@secondoftwo
 \fi
}%
\providecommand \natexlab [1]{#1}%
\providecommand \enquote  [1]{``#1''}%
\providecommand \bibnamefont  [1]{#1}%
\providecommand \bibfnamefont [1]{#1}%
\providecommand \citenamefont [1]{#1}%
\providecommand \href@noop [0]{\@secondoftwo}%
\providecommand \href [0]{\begingroup \@sanitize@url \@href}%
\providecommand \@href[1]{\@@startlink{#1}\@@href}%
\providecommand \@@href[1]{\endgroup#1\@@endlink}%
\providecommand \@sanitize@url [0]{\catcode `\\12\catcode `\$12\catcode `\&12\catcode `\#12\catcode `\^12\catcode `\_12\catcode `\%12\relax}%
\providecommand \@@startlink[1]{}%
\providecommand \@@endlink[0]{}%
\providecommand \url  [0]{\begingroup\@sanitize@url \@url }%
\providecommand \@url [1]{\endgroup\@href {#1}{\urlprefix }}%
\providecommand \urlprefix  [0]{URL }%
\providecommand \Eprint [0]{\href }%
\providecommand \doibase [0]{https://doi.org/}%
\providecommand \selectlanguage [0]{\@gobble}%
\providecommand \bibinfo  [0]{\@secondoftwo}%
\providecommand \bibfield  [0]{\@secondoftwo}%
\providecommand \translation [1]{[#1]}%
\providecommand \BibitemOpen [0]{}%
\providecommand \bibitemStop [0]{}%
\providecommand \bibitemNoStop [0]{.\EOS\space}%
\providecommand \EOS [0]{\spacefactor3000\relax}%
\providecommand \BibitemShut  [1]{\csname bibitem#1\endcsname}%
\let\auto@bib@innerbib\@empty
\bibitem [{\citenamefont {Hasan}\ and\ \citenamefont {Kane}(2010)}]{colloquium}%
  \BibitemOpen
  \bibfield  {author} {\bibinfo {author} {\bibfnamefont {M.~Z.}\ \bibnamefont {Hasan}}\ and\ \bibinfo {author} {\bibfnamefont {C.~L.}\ \bibnamefont {Kane}},\ }\href {https://doi.org/10.1103/RevModPhys.82.3045} {\bibfield  {journal} {\bibinfo  {journal} {Rev. Mod. Phys.}\ }\textbf {\bibinfo {volume} {82}},\ \bibinfo {pages} {3045} (\bibinfo {year} {2010})}\BibitemShut {NoStop}%
\bibitem [{\citenamefont {Kane}\ and\ \citenamefont {Mele}(2005)}]{Kane2005}%
  \BibitemOpen
  \bibfield  {author} {\bibinfo {author} {\bibfnamefont {C.~L.}\ \bibnamefont {Kane}}\ and\ \bibinfo {author} {\bibfnamefont {E.~J.}\ \bibnamefont {Mele}},\ }\href {https://doi.org/10.1103/physrevlett.95.146802} {\bibfield  {journal} {\bibinfo  {journal} {Physical Review Letters}\ }\textbf {\bibinfo {volume} {95}} (\bibinfo {year} {2005}),\ 10.1103/physrevlett.95.146802}\BibitemShut {NoStop}%
\bibitem [{\citenamefont {Qi}\ and\ \citenamefont {Zhang}(2011)}]{t3}%
  \BibitemOpen
  \bibfield  {author} {\bibinfo {author} {\bibfnamefont {X.-L.}\ \bibnamefont {Qi}}\ and\ \bibinfo {author} {\bibfnamefont {S.-C.}\ \bibnamefont {Zhang}},\ }\href {https://doi.org/10.1103/RevModPhys.83.1057} {\bibfield  {journal} {\bibinfo  {journal} {Rev. Mod. Phys.}\ }\textbf {\bibinfo {volume} {83}},\ \bibinfo {pages} {1057} (\bibinfo {year} {2011})}\BibitemShut {NoStop}%
\bibitem [{\citenamefont {Tokura}, \citenamefont {Yasuda},\ and\ \citenamefont {Tsukazaki}(2019)}]{t4}%
  \BibitemOpen
  \bibfield  {author} {\bibinfo {author} {\bibfnamefont {Y.}~\bibnamefont {Tokura}}, \bibinfo {author} {\bibfnamefont {K.}~\bibnamefont {Yasuda}},\ and\ \bibinfo {author} {\bibfnamefont {A.}~\bibnamefont {Tsukazaki}},\ }\href@noop {} {\bibfield  {journal} {\bibinfo  {journal} {Nature Reviews Physics}\ }\textbf {\bibinfo {volume} {1}},\ \bibinfo {pages} {126} (\bibinfo {year} {2019})}\BibitemShut {NoStop}%
\bibitem [{\citenamefont {Moore}(2010)}]{t5}%
  \BibitemOpen
  \bibfield  {author} {\bibinfo {author} {\bibfnamefont {J.~E.}\ \bibnamefont {Moore}},\ }\href@noop {} {\bibfield  {journal} {\bibinfo  {journal} {Nature}\ }\textbf {\bibinfo {volume} {464}},\ \bibinfo {pages} {194} (\bibinfo {year} {2010})}\BibitemShut {NoStop}%
\bibitem [{\citenamefont {Shankar}(2018)}]{t6}%
  \BibitemOpen
  \bibfield  {author} {\bibinfo {author} {\bibfnamefont {R.}~\bibnamefont {Shankar}},\ }\href@noop {} {\bibfield  {journal} {\bibinfo  {journal} {arXiv preprint arXiv:1804.06471}\ } (\bibinfo {year} {2018})}\BibitemShut {NoStop}%
\bibitem [{\citenamefont {Bernevig}, \citenamefont {Hughes},\ and\ \citenamefont {Zhang}(2006)}]{t7}%
  \BibitemOpen
  \bibfield  {author} {\bibinfo {author} {\bibfnamefont {B.~A.}\ \bibnamefont {Bernevig}}, \bibinfo {author} {\bibfnamefont {T.~L.}\ \bibnamefont {Hughes}},\ and\ \bibinfo {author} {\bibfnamefont {S.-C.}\ \bibnamefont {Zhang}},\ }\href@noop {} {\bibfield  {journal} {\bibinfo  {journal} {science}\ }\textbf {\bibinfo {volume} {314}},\ \bibinfo {pages} {1757} (\bibinfo {year} {2006})}\BibitemShut {NoStop}%
\bibitem [{\citenamefont {Fu}, \citenamefont {Kane},\ and\ \citenamefont {Mele}(2007)}]{t8}%
  \BibitemOpen
  \bibfield  {author} {\bibinfo {author} {\bibfnamefont {L.}~\bibnamefont {Fu}}, \bibinfo {author} {\bibfnamefont {C.~L.}\ \bibnamefont {Kane}},\ and\ \bibinfo {author} {\bibfnamefont {E.~J.}\ \bibnamefont {Mele}},\ }\href {https://doi.org/10.1103/PhysRevLett.98.106803} {\bibfield  {journal} {\bibinfo  {journal} {Phys. Rev. Lett.}\ }\textbf {\bibinfo {volume} {98}},\ \bibinfo {pages} {106803} (\bibinfo {year} {2007})}\BibitemShut {NoStop}%
\bibitem [{\citenamefont {Fu}\ and\ \citenamefont {Kane}(2007)}]{strong_weak_TI}%
  \BibitemOpen
  \bibfield  {author} {\bibinfo {author} {\bibfnamefont {L.}~\bibnamefont {Fu}}\ and\ \bibinfo {author} {\bibfnamefont {C.~L.}\ \bibnamefont {Kane}},\ }\href {https://doi.org/10.1103/PhysRevB.76.045302} {\bibfield  {journal} {\bibinfo  {journal} {Phys. Rev. B}\ }\textbf {\bibinfo {volume} {76}},\ \bibinfo {pages} {045302} (\bibinfo {year} {2007})}\BibitemShut {NoStop}%
\bibitem [{\citenamefont {Santos}\ and\ \citenamefont {Gutman}(2015)}]{TR}%
  \BibitemOpen
  \bibfield  {author} {\bibinfo {author} {\bibfnamefont {R.~A.}\ \bibnamefont {Santos}}\ and\ \bibinfo {author} {\bibfnamefont {D.~B.}\ \bibnamefont {Gutman}},\ }\href {https://doi.org/10.1103/PhysRevB.92.075135} {\bibfield  {journal} {\bibinfo  {journal} {Phys. Rev. B}\ }\textbf {\bibinfo {volume} {92}},\ \bibinfo {pages} {075135} (\bibinfo {year} {2015})}\BibitemShut {NoStop}%
\bibitem [{\citenamefont {Zhang}\ \emph {et~al.}(2009)\citenamefont {Zhang}, \citenamefont {Cheng}, \citenamefont {Chen}, \citenamefont {Jia}, \citenamefont {Ma}, \citenamefont {He}, \citenamefont {Wang}, \citenamefont {Zhang}, \citenamefont {Dai}, \citenamefont {Fang}, \citenamefont {Xie},\ and\ \citenamefont {Xue}}]{TR1}%
  \BibitemOpen
  \bibfield  {author} {\bibinfo {author} {\bibfnamefont {T.}~\bibnamefont {Zhang}}, \bibinfo {author} {\bibfnamefont {P.}~\bibnamefont {Cheng}}, \bibinfo {author} {\bibfnamefont {X.}~\bibnamefont {Chen}}, \bibinfo {author} {\bibfnamefont {J.-F.}\ \bibnamefont {Jia}}, \bibinfo {author} {\bibfnamefont {X.}~\bibnamefont {Ma}}, \bibinfo {author} {\bibfnamefont {K.}~\bibnamefont {He}}, \bibinfo {author} {\bibfnamefont {L.}~\bibnamefont {Wang}}, \bibinfo {author} {\bibfnamefont {H.}~\bibnamefont {Zhang}}, \bibinfo {author} {\bibfnamefont {X.}~\bibnamefont {Dai}}, \bibinfo {author} {\bibfnamefont {Z.}~\bibnamefont {Fang}}, \bibinfo {author} {\bibfnamefont {X.}~\bibnamefont {Xie}},\ and\ \bibinfo {author} {\bibfnamefont {Q.-K.}\ \bibnamefont {Xue}},\ }\href {https://doi.org/10.1103/PhysRevLett.103.266803} {\bibfield  {journal} {\bibinfo  {journal} {Phys. Rev. Lett.}\ }\textbf {\bibinfo {volume} {103}},\ \bibinfo {pages} {266803} (\bibinfo {year} {2009})}\BibitemShut {NoStop}%
\bibitem [{\citenamefont {Fu}\ and\ \citenamefont {Kane}(2008)}]{superconducting}%
  \BibitemOpen
  \bibfield  {author} {\bibinfo {author} {\bibfnamefont {L.}~\bibnamefont {Fu}}\ and\ \bibinfo {author} {\bibfnamefont {C.~L.}\ \bibnamefont {Kane}},\ }\href {https://doi.org/10.1103/PhysRevLett.100.096407} {\bibfield  {journal} {\bibinfo  {journal} {Phys. Rev. Lett.}\ }\textbf {\bibinfo {volume} {100}},\ \bibinfo {pages} {096407} (\bibinfo {year} {2008})}\BibitemShut {NoStop}%
\bibitem [{\citenamefont {Fan}\ and\ \citenamefont {Wang}(2016)}]{spintronics}%
  \BibitemOpen
  \bibfield  {author} {\bibinfo {author} {\bibfnamefont {Y.}~\bibnamefont {Fan}}\ and\ \bibinfo {author} {\bibfnamefont {K.~L.}\ \bibnamefont {Wang}},\ }in\ \href@noop {} {\emph {\bibinfo {booktitle} {Spin}}},\ Vol.~\bibinfo {volume} {6}\ (\bibinfo {organization} {World Scientific},\ \bibinfo {year} {2016})\ p.\ \bibinfo {pages} {1640001}\BibitemShut {NoStop}%
\bibitem [{\citenamefont {Pal}\ \emph {et~al.}(2024)\citenamefont {Pal}, \citenamefont {Nandi}, \citenamefont {Nath}, \citenamefont {Pal}, \citenamefont {Sharma}, \citenamefont {Manna}, \citenamefont {Barman},\ and\ \citenamefont {Mitra}}]{pal2024enhancement}%
  \BibitemOpen
  \bibfield  {author} {\bibinfo {author} {\bibfnamefont {S.}~\bibnamefont {Pal}}, \bibinfo {author} {\bibfnamefont {A.}~\bibnamefont {Nandi}}, \bibinfo {author} {\bibfnamefont {S.~G.}\ \bibnamefont {Nath}}, \bibinfo {author} {\bibfnamefont {P.~K.}\ \bibnamefont {Pal}}, \bibinfo {author} {\bibfnamefont {K.}~\bibnamefont {Sharma}}, \bibinfo {author} {\bibfnamefont {S.}~\bibnamefont {Manna}}, \bibinfo {author} {\bibfnamefont {A.}~\bibnamefont {Barman}},\ and\ \bibinfo {author} {\bibfnamefont {C.}~\bibnamefont {Mitra}},\ }\href {https://doi.org/10.1063/5.0192717} {\bibfield  {journal} {\bibinfo  {journal} {Applied Physics Letters}\ }\textbf {\bibinfo {volume} {124}},\ \bibinfo {pages} {112416} (\bibinfo {year} {2024})}\BibitemShut {NoStop}%
\bibitem [{\citenamefont {Nemirovsky}, \citenamefont {Brouk},\ and\ \citenamefont {Jakobson}(2001)}]{low_noise}%
  \BibitemOpen
  \bibfield  {author} {\bibinfo {author} {\bibfnamefont {Y.}~\bibnamefont {Nemirovsky}}, \bibinfo {author} {\bibfnamefont {I.}~\bibnamefont {Brouk}},\ and\ \bibinfo {author} {\bibfnamefont {C.}~\bibnamefont {Jakobson}},\ }\href {https://doi.org/10.1109/16.918240} {\bibfield  {journal} {\bibinfo  {journal} {IEEE Transactions on Electron Devices}\ }\textbf {\bibinfo {volume} {48}},\ \bibinfo {pages} {921} (\bibinfo {year} {2001})}\BibitemShut {NoStop}%
\bibitem [{\citenamefont {Kushwaha}\ \emph {et~al.}(2016)\citenamefont {Kushwaha}, \citenamefont {Pletikosi{\'c}}, \citenamefont {Liang}, \citenamefont {Gyenis}, \citenamefont {Lapidus}, \citenamefont {Tian}, \citenamefont {Zhao}, \citenamefont {Burch}, \citenamefont {Lin}, \citenamefont {Wang} \emph {et~al.}}]{kushwaha2016sn}%
  \BibitemOpen
  \bibfield  {author} {\bibinfo {author} {\bibfnamefont {S.}~\bibnamefont {Kushwaha}}, \bibinfo {author} {\bibfnamefont {I.}~\bibnamefont {Pletikosi{\'c}}}, \bibinfo {author} {\bibfnamefont {T.}~\bibnamefont {Liang}}, \bibinfo {author} {\bibfnamefont {A.}~\bibnamefont {Gyenis}}, \bibinfo {author} {\bibfnamefont {S.}~\bibnamefont {Lapidus}}, \bibinfo {author} {\bibfnamefont {Y.}~\bibnamefont {Tian}}, \bibinfo {author} {\bibfnamefont {H.}~\bibnamefont {Zhao}}, \bibinfo {author} {\bibfnamefont {K.}~\bibnamefont {Burch}}, \bibinfo {author} {\bibfnamefont {J.}~\bibnamefont {Lin}}, \bibinfo {author} {\bibfnamefont {W.}~\bibnamefont {Wang}}, \emph {et~al.},\ }\href {https://doi.org/10.1038/ncomms11456} {\bibfield  {journal} {\bibinfo  {journal} {Nature communications}\ }\textbf {\bibinfo {volume} {7}},\ \bibinfo {pages} {11456} (\bibinfo {year} {2016})}\BibitemShut {NoStop}%
\bibitem [{\citenamefont {German}\ \emph {et~al.}(2019)\citenamefont {German}, \citenamefont {Komleva}, \citenamefont {Stein}, \citenamefont {Mazurenko}, \citenamefont {Wang}, \citenamefont {Streltsov}, \citenamefont {Ando},\ and\ \citenamefont {van Loosdrecht}}]{PhysRevMaterials.3.054204}%
  \BibitemOpen
  \bibfield  {author} {\bibinfo {author} {\bibfnamefont {R.}~\bibnamefont {German}}, \bibinfo {author} {\bibfnamefont {E.~V.}\ \bibnamefont {Komleva}}, \bibinfo {author} {\bibfnamefont {P.}~\bibnamefont {Stein}}, \bibinfo {author} {\bibfnamefont {V.~G.}\ \bibnamefont {Mazurenko}}, \bibinfo {author} {\bibfnamefont {Z.}~\bibnamefont {Wang}}, \bibinfo {author} {\bibfnamefont {S.~V.}\ \bibnamefont {Streltsov}}, \bibinfo {author} {\bibfnamefont {Y.}~\bibnamefont {Ando}},\ and\ \bibinfo {author} {\bibfnamefont {P.~H.~M.}\ \bibnamefont {van Loosdrecht}},\ }\href {https://doi.org/10.1103/PhysRevMaterials.3.054204} {\bibfield  {journal} {\bibinfo  {journal} {Phys. Rev. Mater.}\ }\textbf {\bibinfo {volume} {3}},\ \bibinfo {pages} {054204} (\bibinfo {year} {2019})}\BibitemShut {NoStop}%
\bibitem [{\citenamefont {Xu}\ and\ \citenamefont {Chen}(2021)}]{flakes}%
  \BibitemOpen
  \bibfield  {author} {\bibinfo {author} {\bibfnamefont {Y.}~\bibnamefont {Xu}}\ and\ \bibinfo {author} {\bibfnamefont {Y.~P.}\ \bibnamefont {Chen}},\ }in\ \href {https://doi.org/https://doi.org/10.1016/bs.semsem.2021.07.002} {\emph {\bibinfo {booktitle} {Topological Insulator and Related Topics}}},\ \bibinfo {series} {Semiconductors and Semimetals}, Vol.\ \bibinfo {volume} {108},\ \bibinfo {editor} {edited by\ \bibinfo {editor} {\bibfnamefont {L.}~\bibnamefont {Li}}\ and\ \bibinfo {editor} {\bibfnamefont {K.}~\bibnamefont {Sun}}}\ (\bibinfo  {publisher} {Elsevier},\ \bibinfo {year} {2021})\ pp.\ \bibinfo {pages} {73--124}\BibitemShut {NoStop}%
\bibitem [{\citenamefont {Pandey}\ \emph {et~al.}(2019)\citenamefont {Pandey}, \citenamefont {Singh}, \citenamefont {Ghosh}, \citenamefont {Manna}, \citenamefont {Gopal},\ and\ \citenamefont {Mitra}}]{pulsed}%
  \BibitemOpen
  \bibfield  {author} {\bibinfo {author} {\bibfnamefont {A.}~\bibnamefont {Pandey}}, \bibinfo {author} {\bibfnamefont {S.}~\bibnamefont {Singh}}, \bibinfo {author} {\bibfnamefont {B.}~\bibnamefont {Ghosh}}, \bibinfo {author} {\bibfnamefont {S.}~\bibnamefont {Manna}}, \bibinfo {author} {\bibfnamefont {R.}~\bibnamefont {Gopal}},\ and\ \bibinfo {author} {\bibfnamefont {C.}~\bibnamefont {Mitra}},\ }\href@noop {} {\enquote {\bibinfo {title} {Pulsed laser deposition of highly c-axis oriented thin films of bsts topological insulator},}\ } (\bibinfo {year} {2019}),\ \Eprint {https://arxiv.org/abs/1910.08100} {arXiv:1910.08100 [cond-mat.mtrl-sci]} \BibitemShut {NoStop}%
\bibitem [{\citenamefont {Singh}\ \emph {et~al.}(2017)\citenamefont {Singh}, \citenamefont {Gopal}, \citenamefont {Sarkar}, \citenamefont {Pandey}, \citenamefont {Patel},\ and\ \citenamefont {Mitra}}]{pld2}%
  \BibitemOpen
  \bibfield  {author} {\bibinfo {author} {\bibfnamefont {S.}~\bibnamefont {Singh}}, \bibinfo {author} {\bibfnamefont {R.}~\bibnamefont {Gopal}}, \bibinfo {author} {\bibfnamefont {J.}~\bibnamefont {Sarkar}}, \bibinfo {author} {\bibfnamefont {A.}~\bibnamefont {Pandey}}, \bibinfo {author} {\bibfnamefont {B.~G.}\ \bibnamefont {Patel}},\ and\ \bibinfo {author} {\bibfnamefont {C.}~\bibnamefont {Mitra}},\ }\href@noop {} {\bibfield  {journal} {\bibinfo  {journal} {Journal of Physics: Condensed Matter}\ }\textbf {\bibinfo {volume} {29}},\ \bibinfo {pages} {505601} (\bibinfo {year} {2017})}\BibitemShut {NoStop}%
\bibitem [{\citenamefont {Mitra}\ \emph {et~al.}(2002)\citenamefont {Mitra}, \citenamefont {Köbernik}, \citenamefont {Dörr}, \citenamefont {Müller}, \citenamefont {Schultz}, \citenamefont {Raychaudhuri}, \citenamefont {Pinto},\ and\ \citenamefont {Wieser}}]{mitra2002magnetotransport}%
  \BibitemOpen
  \bibfield  {author} {\bibinfo {author} {\bibfnamefont {C.}~\bibnamefont {Mitra}}, \bibinfo {author} {\bibfnamefont {G.}~\bibnamefont {Köbernik}}, \bibinfo {author} {\bibfnamefont {K.}~\bibnamefont {Dörr}}, \bibinfo {author} {\bibfnamefont {K.-H.}\ \bibnamefont {Müller}}, \bibinfo {author} {\bibfnamefont {L.}~\bibnamefont {Schultz}}, \bibinfo {author} {\bibfnamefont {P.}~\bibnamefont {Raychaudhuri}}, \bibinfo {author} {\bibfnamefont {R.}~\bibnamefont {Pinto}},\ and\ \bibinfo {author} {\bibfnamefont {E.}~\bibnamefont {Wieser}},\ }\href {https://doi.org/10.1063/1.1451842} {\bibfield  {journal} {\bibinfo  {journal} {Journal of Applied Physics}\ }\textbf {\bibinfo {volume} {91}},\ \bibinfo {pages} {7715} (\bibinfo {year} {2002})}\BibitemShut {NoStop}%
\bibitem [{\citenamefont {Bose}\ \emph {et~al.}(2006)\citenamefont {Bose}, \citenamefont {Banerjee}, \citenamefont {Genc}, \citenamefont {Raychaudhuri}, \citenamefont {Fraser},\ and\ \citenamefont {Ayyub}}]{granular}%
  \BibitemOpen
  \bibfield  {author} {\bibinfo {author} {\bibfnamefont {S.}~\bibnamefont {Bose}}, \bibinfo {author} {\bibfnamefont {R.}~\bibnamefont {Banerjee}}, \bibinfo {author} {\bibfnamefont {A.}~\bibnamefont {Genc}}, \bibinfo {author} {\bibfnamefont {P.}~\bibnamefont {Raychaudhuri}}, \bibinfo {author} {\bibfnamefont {H.~L.}\ \bibnamefont {Fraser}},\ and\ \bibinfo {author} {\bibfnamefont {P.}~\bibnamefont {Ayyub}},\ }\href {https://doi.org/10.1088/0953-8984/18/19/010} {\bibfield  {journal} {\bibinfo  {journal} {Journal of Physics: Condensed Matter}\ }\textbf {\bibinfo {volume} {18}},\ \bibinfo {pages} {4553} (\bibinfo {year} {2006})}\BibitemShut {NoStop}%
\bibitem [{\citenamefont {Shtern}\ \emph {et~al.}(2021)\citenamefont {Shtern}, \citenamefont {Rogachev}, \citenamefont {Shtern}, \citenamefont {Sherchenkov}, \citenamefont {Babich}, \citenamefont {Korchagin},\ and\ \citenamefont {Nikulin}}]{thermoelectric}%
  \BibitemOpen
  \bibfield  {author} {\bibinfo {author} {\bibfnamefont {M.}~\bibnamefont {Shtern}}, \bibinfo {author} {\bibfnamefont {M.}~\bibnamefont {Rogachev}}, \bibinfo {author} {\bibfnamefont {Y.}~\bibnamefont {Shtern}}, \bibinfo {author} {\bibfnamefont {A.}~\bibnamefont {Sherchenkov}}, \bibinfo {author} {\bibfnamefont {A.}~\bibnamefont {Babich}}, \bibinfo {author} {\bibfnamefont {E.}~\bibnamefont {Korchagin}},\ and\ \bibinfo {author} {\bibfnamefont {D.}~\bibnamefont {Nikulin}},\ }\href@noop {} {\bibfield  {journal} {\bibinfo  {journal} {Journal of Alloys and Compounds}\ }\textbf {\bibinfo {volume} {877}},\ \bibinfo {pages} {160328} (\bibinfo {year} {2021})}\BibitemShut {NoStop}%
\bibitem [{\citenamefont {Hooge}, \citenamefont {Kleinpenning},\ and\ \citenamefont {Vandamme}(1981{\natexlab{a}})}]{hooge}%
  \BibitemOpen
  \bibfield  {author} {\bibinfo {author} {\bibfnamefont {F.}~\bibnamefont {Hooge}}, \bibinfo {author} {\bibfnamefont {T.}~\bibnamefont {Kleinpenning}},\ and\ \bibinfo {author} {\bibfnamefont {L.~K.}\ \bibnamefont {Vandamme}},\ }\href@noop {} {\bibfield  {journal} {\bibinfo  {journal} {Reports on progress in Physics}\ }\textbf {\bibinfo {volume} {44}},\ \bibinfo {pages} {479} (\bibinfo {year} {1981}{\natexlab{a}})}\BibitemShut {NoStop}%
\bibitem [{\citenamefont {Milotti}(2002)}]{n2}%
  \BibitemOpen
  \bibfield  {author} {\bibinfo {author} {\bibfnamefont {E.}~\bibnamefont {Milotti}},\ }\href@noop {} {\bibfield  {journal} {\bibinfo  {journal} {arXiv preprint physics/0204033}\ } (\bibinfo {year} {2002})}\BibitemShut {NoStop}%
\bibitem [{\citenamefont {Hooge}, \citenamefont {Kleinpenning},\ and\ \citenamefont {Vandamme}(1981{\natexlab{b}})}]{n3}%
  \BibitemOpen
  \bibfield  {author} {\bibinfo {author} {\bibfnamefont {F.}~\bibnamefont {Hooge}}, \bibinfo {author} {\bibfnamefont {T.}~\bibnamefont {Kleinpenning}},\ and\ \bibinfo {author} {\bibfnamefont {L.~K.}\ \bibnamefont {Vandamme}},\ }\href@noop {} {\bibfield  {journal} {\bibinfo  {journal} {Reports on progress in Physics}\ }\textbf {\bibinfo {volume} {44}},\ \bibinfo {pages} {479} (\bibinfo {year} {1981}{\natexlab{b}})}\BibitemShut {NoStop}%
\bibitem [{\citenamefont {Dutta}\ and\ \citenamefont {Horn}(1981{\natexlab{a}})}]{n4}%
  \BibitemOpen
  \bibfield  {author} {\bibinfo {author} {\bibfnamefont {P.}~\bibnamefont {Dutta}}\ and\ \bibinfo {author} {\bibfnamefont {P.~M.}\ \bibnamefont {Horn}},\ }\href {https://doi.org/10.1103/RevModPhys.53.497} {\bibfield  {journal} {\bibinfo  {journal} {Rev. Mod. Phys.}\ }\textbf {\bibinfo {volume} {53}},\ \bibinfo {pages} {497} (\bibinfo {year} {1981}{\natexlab{a}})}\BibitemShut {NoStop}%
\bibitem [{\citenamefont {Islam}, \citenamefont {Shamim},\ and\ \citenamefont {Ghosh}(2023)}]{islam2023benchmarking}%
  \BibitemOpen
  \bibfield  {author} {\bibinfo {author} {\bibfnamefont {S.}~\bibnamefont {Islam}}, \bibinfo {author} {\bibfnamefont {S.}~\bibnamefont {Shamim}},\ and\ \bibinfo {author} {\bibfnamefont {A.}~\bibnamefont {Ghosh}},\ }\href@noop {} {\bibfield  {journal} {\bibinfo  {journal} {Advanced Materials}\ }\textbf {\bibinfo {volume} {35}},\ \bibinfo {pages} {2109671} (\bibinfo {year} {2023})}\BibitemShut {NoStop}%
\bibitem [{\citenamefont {Pal}\ \emph {et~al.}(2011)\citenamefont {Pal}, \citenamefont {Ghatak}, \citenamefont {Kochat}, \citenamefont {Sneha}, \citenamefont {Sampathkumar}, \citenamefont {Raghavan},\ and\ \citenamefont {Ghosh}}]{pal2011microscopic}%
  \BibitemOpen
  \bibfield  {author} {\bibinfo {author} {\bibfnamefont {A.~N.}\ \bibnamefont {Pal}}, \bibinfo {author} {\bibfnamefont {S.}~\bibnamefont {Ghatak}}, \bibinfo {author} {\bibfnamefont {V.}~\bibnamefont {Kochat}}, \bibinfo {author} {\bibfnamefont {E.~S.}\ \bibnamefont {Sneha}}, \bibinfo {author} {\bibfnamefont {A.}~\bibnamefont {Sampathkumar}}, \bibinfo {author} {\bibfnamefont {S.}~\bibnamefont {Raghavan}},\ and\ \bibinfo {author} {\bibfnamefont {A.}~\bibnamefont {Ghosh}},\ }\href {https://doi.org/10.1021/nn103273n} {\bibfield  {journal} {\bibinfo  {journal} {ACS Nano}\ }\textbf {\bibinfo {volume} {5}},\ \bibinfo {pages} {2075} (\bibinfo {year} {2011})},\ \bibinfo {note} {pMID: 21332148},\ \Eprint {https://arxiv.org/abs/https://doi.org/10.1021/nn103273n} {https://doi.org/10.1021/nn103273n} \BibitemShut {NoStop}%
\bibitem [{\citenamefont {Balandin}(2013)}]{balandin2013low}%
  \BibitemOpen
  \bibfield  {author} {\bibinfo {author} {\bibfnamefont {A.~A.}\ \bibnamefont {Balandin}},\ }\href@noop {} {\bibfield  {journal} {\bibinfo  {journal} {Nature nanotechnology}\ }\textbf {\bibinfo {volume} {8}},\ \bibinfo {pages} {549} (\bibinfo {year} {2013})}\BibitemShut {NoStop}%
\bibitem [{\citenamefont {Rehman}\ \emph {et~al.}(2023)\citenamefont {Rehman}, \citenamefont {Cywinski}, \citenamefont {Knap}, \citenamefont {Smulko}, \citenamefont {Balandin},\ and\ \citenamefont {Rumyantsev}}]{rehman2023low}%
  \BibitemOpen
  \bibfield  {author} {\bibinfo {author} {\bibfnamefont {A.}~\bibnamefont {Rehman}}, \bibinfo {author} {\bibfnamefont {G.}~\bibnamefont {Cywinski}}, \bibinfo {author} {\bibfnamefont {W.}~\bibnamefont {Knap}}, \bibinfo {author} {\bibfnamefont {J.}~\bibnamefont {Smulko}}, \bibinfo {author} {\bibfnamefont {A.~A.}\ \bibnamefont {Balandin}},\ and\ \bibinfo {author} {\bibfnamefont {S.}~\bibnamefont {Rumyantsev}},\ }\href {https://doi.org/10.1063/5.0143641} {\bibfield  {journal} {\bibinfo  {journal} {Applied Physics Letters}\ }\textbf {\bibinfo {volume} {122}},\ \bibinfo {pages} {090602} (\bibinfo {year} {2023})}\BibitemShut {NoStop}%
\bibitem [{\citenamefont {Scanlon}\ \emph {et~al.}(2012)\citenamefont {Scanlon}, \citenamefont {King}, \citenamefont {Singh}, \citenamefont {de~la Torre}, \citenamefont {Walker}, \citenamefont {Balakrishnan}, \citenamefont {Baumberger},\ and\ \citenamefont {Catlow}}]{antisite}%
  \BibitemOpen
  \bibfield  {author} {\bibinfo {author} {\bibfnamefont {D.~O.}\ \bibnamefont {Scanlon}}, \bibinfo {author} {\bibfnamefont {P.~D.~C.}\ \bibnamefont {King}}, \bibinfo {author} {\bibfnamefont {R.~P.}\ \bibnamefont {Singh}}, \bibinfo {author} {\bibfnamefont {A.}~\bibnamefont {de~la Torre}}, \bibinfo {author} {\bibfnamefont {S.~M.}\ \bibnamefont {Walker}}, \bibinfo {author} {\bibfnamefont {G.}~\bibnamefont {Balakrishnan}}, \bibinfo {author} {\bibfnamefont {F.}~\bibnamefont {Baumberger}},\ and\ \bibinfo {author} {\bibfnamefont {C.~R.~A.}\ \bibnamefont {Catlow}},\ }\href {https://doi.org/10.1002/adma.201200187} {\bibfield  {journal} {\bibinfo  {journal} {Advanced Materials}\ }\textbf {\bibinfo {volume} {24}},\ \bibinfo {pages} {2154–2158} (\bibinfo {year} {2012})}\BibitemShut {NoStop}%
\bibitem [{\citenamefont {Netsou}\ \emph {et~al.}(2020)\citenamefont {Netsou}, \citenamefont {Muzychenko}, \citenamefont {Dausy}, \citenamefont {Chen}, \citenamefont {Song}, \citenamefont {Schouteden}, \citenamefont {Van~Bael},\ and\ \citenamefont {Van~Haesendonck}}]{antisite2}%
  \BibitemOpen
  \bibfield  {author} {\bibinfo {author} {\bibfnamefont {A.-M.}\ \bibnamefont {Netsou}}, \bibinfo {author} {\bibfnamefont {D.~A.}\ \bibnamefont {Muzychenko}}, \bibinfo {author} {\bibfnamefont {H.}~\bibnamefont {Dausy}}, \bibinfo {author} {\bibfnamefont {T.}~\bibnamefont {Chen}}, \bibinfo {author} {\bibfnamefont {F.}~\bibnamefont {Song}}, \bibinfo {author} {\bibfnamefont {K.}~\bibnamefont {Schouteden}}, \bibinfo {author} {\bibfnamefont {M.~J.}\ \bibnamefont {Van~Bael}},\ and\ \bibinfo {author} {\bibfnamefont {C.}~\bibnamefont {Van~Haesendonck}},\ }\href@noop {} {\bibfield  {journal} {\bibinfo  {journal} {ACS nano}\ }\textbf {\bibinfo {volume} {14}},\ \bibinfo {pages} {13172} (\bibinfo {year} {2020})}\BibitemShut {NoStop}%
\bibitem [{\citenamefont {Tumelero}, \citenamefont {Faccio},\ and\ \citenamefont {Pasa}(2016)}]{antisite3}%
  \BibitemOpen
  \bibfield  {author} {\bibinfo {author} {\bibfnamefont {M.~A.}\ \bibnamefont {Tumelero}}, \bibinfo {author} {\bibfnamefont {R.}~\bibnamefont {Faccio}},\ and\ \bibinfo {author} {\bibfnamefont {A.~A.}\ \bibnamefont {Pasa}},\ }\href@noop {} {\bibfield  {journal} {\bibinfo  {journal} {Journal of Physics: Condensed Matter}\ }\textbf {\bibinfo {volume} {28}},\ \bibinfo {pages} {425801} (\bibinfo {year} {2016})}\BibitemShut {NoStop}%
\bibitem [{\citenamefont {Skinner}, \citenamefont {Chen},\ and\ \citenamefont {Shklovskii}(2012)}]{skinner2012bulk}%
  \BibitemOpen
  \bibfield  {author} {\bibinfo {author} {\bibfnamefont {B.}~\bibnamefont {Skinner}}, \bibinfo {author} {\bibfnamefont {T.}~\bibnamefont {Chen}},\ and\ \bibinfo {author} {\bibfnamefont {B.~I.}\ \bibnamefont {Shklovskii}},\ }\href {https://doi.org/10.1103/PhysRevLett.109.176801} {\bibfield  {journal} {\bibinfo  {journal} {Phys. Rev. Lett.}\ }\textbf {\bibinfo {volume} {109}},\ \bibinfo {pages} {176801} (\bibinfo {year} {2012})}\BibitemShut {NoStop}%
\bibitem [{\citenamefont {Knispel}\ \emph {et~al.}(2017)\citenamefont {Knispel}, \citenamefont {Jolie}, \citenamefont {Borgwardt}, \citenamefont {Lux}, \citenamefont {Wang}, \citenamefont {Ando}, \citenamefont {Rosch}, \citenamefont {Michely},\ and\ \citenamefont {Gr\"uninger}}]{PhysRevB.96.195135}%
  \BibitemOpen
  \bibfield  {author} {\bibinfo {author} {\bibfnamefont {T.}~\bibnamefont {Knispel}}, \bibinfo {author} {\bibfnamefont {W.}~\bibnamefont {Jolie}}, \bibinfo {author} {\bibfnamefont {N.}~\bibnamefont {Borgwardt}}, \bibinfo {author} {\bibfnamefont {J.}~\bibnamefont {Lux}}, \bibinfo {author} {\bibfnamefont {Z.}~\bibnamefont {Wang}}, \bibinfo {author} {\bibfnamefont {Y.}~\bibnamefont {Ando}}, \bibinfo {author} {\bibfnamefont {A.}~\bibnamefont {Rosch}}, \bibinfo {author} {\bibfnamefont {T.}~\bibnamefont {Michely}},\ and\ \bibinfo {author} {\bibfnamefont {M.}~\bibnamefont {Gr\"uninger}},\ }\href {https://doi.org/10.1103/PhysRevB.96.195135} {\bibfield  {journal} {\bibinfo  {journal} {Phys. Rev. B}\ }\textbf {\bibinfo {volume} {96}},\ \bibinfo {pages} {195135} (\bibinfo {year} {2017})}\BibitemShut {NoStop}%
\bibitem [{\citenamefont {Islam}\ \emph {et~al.}(2017)\citenamefont {Islam}, \citenamefont {Bhattacharyya}, \citenamefont {Kandala}, \citenamefont {Richardella}, \citenamefont {Samarth},\ and\ \citenamefont {Ghosh}}]{islam2017bulk}%
  \BibitemOpen
  \bibfield  {author} {\bibinfo {author} {\bibfnamefont {S.}~\bibnamefont {Islam}}, \bibinfo {author} {\bibfnamefont {S.}~\bibnamefont {Bhattacharyya}}, \bibinfo {author} {\bibfnamefont {A.}~\bibnamefont {Kandala}}, \bibinfo {author} {\bibfnamefont {A.}~\bibnamefont {Richardella}}, \bibinfo {author} {\bibfnamefont {N.}~\bibnamefont {Samarth}},\ and\ \bibinfo {author} {\bibfnamefont {A.}~\bibnamefont {Ghosh}},\ }\href {https://doi.org/10.1103/PhysRevLett.109.176801} {\bibfield  {journal} {\bibinfo  {journal} {Applied Physics Letters}\ }\textbf {\bibinfo {volume} {111}} (\bibinfo {year} {2017}),\ 10.1103/PhysRevLett.109.176801}\BibitemShut {NoStop}%
\bibitem [{\citenamefont {Brahlek}\ \emph {et~al.}(2012)\citenamefont {Brahlek}, \citenamefont {Bansal}, \citenamefont {Koirala}, \citenamefont {Xu}, \citenamefont {Neupane}, \citenamefont {Liu}, \citenamefont {Hasan},\ and\ \citenamefont {Oh}}]{brahlek1}%
  \BibitemOpen
  \bibfield  {author} {\bibinfo {author} {\bibfnamefont {M.}~\bibnamefont {Brahlek}}, \bibinfo {author} {\bibfnamefont {N.}~\bibnamefont {Bansal}}, \bibinfo {author} {\bibfnamefont {N.}~\bibnamefont {Koirala}}, \bibinfo {author} {\bibfnamefont {S.-Y.}\ \bibnamefont {Xu}}, \bibinfo {author} {\bibfnamefont {M.}~\bibnamefont {Neupane}}, \bibinfo {author} {\bibfnamefont {C.}~\bibnamefont {Liu}}, \bibinfo {author} {\bibfnamefont {M.~Z.}\ \bibnamefont {Hasan}},\ and\ \bibinfo {author} {\bibfnamefont {S.}~\bibnamefont {Oh}},\ }\href {https://doi.org/10.1103/PhysRevLett.109.186403} {\bibfield  {journal} {\bibinfo  {journal} {Phys. Rev. Lett.}\ }\textbf {\bibinfo {volume} {109}},\ \bibinfo {pages} {186403} (\bibinfo {year} {2012})}\BibitemShut {NoStop}%
\bibitem [{\citenamefont {Wu}\ \emph {et~al.}(2013)\citenamefont {Wu}, \citenamefont {Brahlek}, \citenamefont {Vald{\'e}s~Aguilar}, \citenamefont {Stier}, \citenamefont {Morris}, \citenamefont {Lubashevsky}, \citenamefont {Bilbro}, \citenamefont {Bansal}, \citenamefont {Oh},\ and\ \citenamefont {Armitage}}]{brahlek2}%
  \BibitemOpen
  \bibfield  {author} {\bibinfo {author} {\bibfnamefont {L.}~\bibnamefont {Wu}}, \bibinfo {author} {\bibfnamefont {M.}~\bibnamefont {Brahlek}}, \bibinfo {author} {\bibfnamefont {R.}~\bibnamefont {Vald{\'e}s~Aguilar}}, \bibinfo {author} {\bibfnamefont {A.}~\bibnamefont {Stier}}, \bibinfo {author} {\bibfnamefont {C.}~\bibnamefont {Morris}}, \bibinfo {author} {\bibfnamefont {Y.}~\bibnamefont {Lubashevsky}}, \bibinfo {author} {\bibfnamefont {L.}~\bibnamefont {Bilbro}}, \bibinfo {author} {\bibfnamefont {N.}~\bibnamefont {Bansal}}, \bibinfo {author} {\bibfnamefont {S.}~\bibnamefont {Oh}},\ and\ \bibinfo {author} {\bibfnamefont {N.}~\bibnamefont {Armitage}},\ }\href@noop {} {\bibfield  {journal} {\bibinfo  {journal} {Nature Physics}\ }\textbf {\bibinfo {volume} {9}},\ \bibinfo {pages} {410} (\bibinfo {year} {2013})}\BibitemShut {NoStop}%
\bibitem [{\citenamefont {Hegde}\ \emph {et~al.}(2020)\citenamefont {Hegde}, \citenamefont {Prabhu}, \citenamefont {Rao},\ and\ \citenamefont {Babu}}]{transport_In}%
  \BibitemOpen
  \bibfield  {author} {\bibinfo {author} {\bibfnamefont {G.~S.}\ \bibnamefont {Hegde}}, \bibinfo {author} {\bibfnamefont {A.}~\bibnamefont {Prabhu}}, \bibinfo {author} {\bibfnamefont {A.}~\bibnamefont {Rao}},\ and\ \bibinfo {author} {\bibfnamefont {P.}~\bibnamefont {Babu}},\ }\href@noop {} {\bibfield  {journal} {\bibinfo  {journal} {Physica B: Condensed Matter}\ }\textbf {\bibinfo {volume} {584}},\ \bibinfo {pages} {412087} (\bibinfo {year} {2020})}\BibitemShut {NoStop}%
\bibitem [{\citenamefont {Syers}\ \emph {et~al.}(2015)\citenamefont {Syers}, \citenamefont {Kim}, \citenamefont {Fuhrer},\ and\ \citenamefont {Paglione}}]{Syer_fit}%
  \BibitemOpen
  \bibfield  {author} {\bibinfo {author} {\bibfnamefont {P.}~\bibnamefont {Syers}}, \bibinfo {author} {\bibfnamefont {D.}~\bibnamefont {Kim}}, \bibinfo {author} {\bibfnamefont {M.~S.}\ \bibnamefont {Fuhrer}},\ and\ \bibinfo {author} {\bibfnamefont {J.}~\bibnamefont {Paglione}},\ }\href {https://doi.org/10.1103/PhysRevLett.114.096601} {\bibfield  {journal} {\bibinfo  {journal} {Phys. Rev. Lett.}\ }\textbf {\bibinfo {volume} {114}},\ \bibinfo {pages} {096601} (\bibinfo {year} {2015})}\BibitemShut {NoStop}%
\bibitem [{\citenamefont {Zhong}\ \emph {et~al.}(2015)\citenamefont {Zhong}, \citenamefont {He}, \citenamefont {Schneeloch}, \citenamefont {Zhang}, \citenamefont {Liu}, \citenamefont {Pletikosi\ifmmode~\acute{c}\else \'{c}\fi{}}, \citenamefont {Yilmaz}, \citenamefont {Sinkovic}, \citenamefont {Li}, \citenamefont {Ku}, \citenamefont {Valla}, \citenamefont {Tranquada},\ and\ \citenamefont {Gu}}]{PbSnTe_fit}%
  \BibitemOpen
  \bibfield  {author} {\bibinfo {author} {\bibfnamefont {R.}~\bibnamefont {Zhong}}, \bibinfo {author} {\bibfnamefont {X.}~\bibnamefont {He}}, \bibinfo {author} {\bibfnamefont {J.~A.}\ \bibnamefont {Schneeloch}}, \bibinfo {author} {\bibfnamefont {C.}~\bibnamefont {Zhang}}, \bibinfo {author} {\bibfnamefont {T.}~\bibnamefont {Liu}}, \bibinfo {author} {\bibfnamefont {I.}~\bibnamefont {Pletikosi\ifmmode~\acute{c}\else \'{c}\fi{}}}, \bibinfo {author} {\bibfnamefont {T.}~\bibnamefont {Yilmaz}}, \bibinfo {author} {\bibfnamefont {B.}~\bibnamefont {Sinkovic}}, \bibinfo {author} {\bibfnamefont {Q.}~\bibnamefont {Li}}, \bibinfo {author} {\bibfnamefont {W.}~\bibnamefont {Ku}}, \bibinfo {author} {\bibfnamefont {T.}~\bibnamefont {Valla}}, \bibinfo {author} {\bibfnamefont {J.~M.}\ \bibnamefont {Tranquada}},\ and\ \bibinfo {author} {\bibfnamefont {G.}~\bibnamefont {Gu}},\ }\href {https://doi.org/10.1103/PhysRevB.91.195321} {\bibfield  {journal} {\bibinfo  {journal} {Phys. Rev. B}\ }\textbf {\bibinfo {volume} {91}},\ \bibinfo
  {pages} {195321} (\bibinfo {year} {2015})}\BibitemShut {NoStop}%
\bibitem [{\citenamefont {Mitra}\ \emph {et~al.}(2001)\citenamefont {Mitra}, \citenamefont {Raychaudhuri}, \citenamefont {John}, \citenamefont {Dhar}, \citenamefont {Nigam},\ and\ \citenamefont {Pinto}}]{mitra2001growth}%
  \BibitemOpen
  \bibfield  {author} {\bibinfo {author} {\bibfnamefont {C.}~\bibnamefont {Mitra}}, \bibinfo {author} {\bibfnamefont {P.}~\bibnamefont {Raychaudhuri}}, \bibinfo {author} {\bibfnamefont {J.}~\bibnamefont {John}}, \bibinfo {author} {\bibfnamefont {S.}~\bibnamefont {Dhar}}, \bibinfo {author} {\bibfnamefont {A.}~\bibnamefont {Nigam}},\ and\ \bibinfo {author} {\bibfnamefont {R.}~\bibnamefont {Pinto}},\ }\href@noop {} {\bibfield  {journal} {\bibinfo  {journal} {Journal of Applied Physics}\ }\textbf {\bibinfo {volume} {89}},\ \bibinfo {pages} {524} (\bibinfo {year} {2001})}\BibitemShut {NoStop}%
\bibitem [{\citenamefont {Lošťák}\ \emph {et~al.}(1993)\citenamefont {Lošťák}, \citenamefont {Navrátil}, \citenamefont {Šrámkova},\ and\ \citenamefont {Horák}}]{suppression}%
  \BibitemOpen
  \bibfield  {author} {\bibinfo {author} {\bibfnamefont {P.}~\bibnamefont {Lošťák}}, \bibinfo {author} {\bibfnamefont {J.}~\bibnamefont {Navrátil}}, \bibinfo {author} {\bibfnamefont {J.}~\bibnamefont {Šrámkova}},\ and\ \bibinfo {author} {\bibfnamefont {J.}~\bibnamefont {Horák}},\ }\href {https://doi.org/https://doi.org/10.1002/pssa.2211350217} {\bibfield  {journal} {\bibinfo  {journal} {physica status solidi (a)}\ }\textbf {\bibinfo {volume} {135}},\ \bibinfo {pages} {519} (\bibinfo {year} {1993})},\ \Eprint {https://arxiv.org/abs/https://onlinelibrary.wiley.com/doi/pdf/10.1002/pssa.2211350217} {https://onlinelibrary.wiley.com/doi/pdf/10.1002/pssa.2211350217} \BibitemShut {NoStop}%
\bibitem [{\citenamefont {Dutta}\ and\ \citenamefont {Horn}(1981{\natexlab{b}})}]{ohm_law}%
  \BibitemOpen
  \bibfield  {author} {\bibinfo {author} {\bibfnamefont {P.}~\bibnamefont {Dutta}}\ and\ \bibinfo {author} {\bibfnamefont {P.}~\bibnamefont {Horn}},\ }\href@noop {} {\bibfield  {journal} {\bibinfo  {journal} {Reviews of Modern physics}\ }\textbf {\bibinfo {volume} {53}},\ \bibinfo {pages} {497} (\bibinfo {year} {1981}{\natexlab{b}})}\BibitemShut {NoStop}%
\bibitem [{\citenamefont {Biswas}\ \emph {et~al.}(2019)\citenamefont {Biswas}, \citenamefont {Gopal}, \citenamefont {Singh}, \citenamefont {Kant}, \citenamefont {Mitra},\ and\ \citenamefont {Bid}}]{biswas2019resistance}%
  \BibitemOpen
  \bibfield  {author} {\bibinfo {author} {\bibfnamefont {S.}~\bibnamefont {Biswas}}, \bibinfo {author} {\bibfnamefont {R.~K.}\ \bibnamefont {Gopal}}, \bibinfo {author} {\bibfnamefont {S.}~\bibnamefont {Singh}}, \bibinfo {author} {\bibfnamefont {R.}~\bibnamefont {Kant}}, \bibinfo {author} {\bibfnamefont {C.}~\bibnamefont {Mitra}},\ and\ \bibinfo {author} {\bibfnamefont {A.}~\bibnamefont {Bid}},\ }\href {https://doi.org/10.1063/1.5119288} {\bibfield  {journal} {\bibinfo  {journal} {Applied Physics Letters}\ }\textbf {\bibinfo {volume} {115}},\ \bibinfo {pages} {131601} (\bibinfo {year} {2019})}\BibitemShut {NoStop}%
\bibitem [{\citenamefont {Grassi}, \citenamefont {Colombo},\ and\ \citenamefont {Camin}(2001)}]{freq_shift}%
  \BibitemOpen
  \bibfield  {author} {\bibinfo {author} {\bibfnamefont {V.}~\bibnamefont {Grassi}}, \bibinfo {author} {\bibfnamefont {C.}~\bibnamefont {Colombo}},\ and\ \bibinfo {author} {\bibfnamefont {D.}~\bibnamefont {Camin}},\ }\href {https://doi.org/10.1109/16.974725} {\bibfield  {journal} {\bibinfo  {journal} {IEEE Transactions on Electron Devices}\ }\textbf {\bibinfo {volume} {48}},\ \bibinfo {pages} {2899} (\bibinfo {year} {2001})}\BibitemShut {NoStop}%
\bibitem [{\citenamefont {Bhattacharyya}\ \emph {et~al.}(2015)\citenamefont {Bhattacharyya}, \citenamefont {Banerjee}, \citenamefont {Nhalil}, \citenamefont {Islam}, \citenamefont {Dasgupta}, \citenamefont {Elizabeth},\ and\ \citenamefont {Ghosh}}]{110K_peak}%
  \BibitemOpen
  \bibfield  {author} {\bibinfo {author} {\bibfnamefont {S.}~\bibnamefont {Bhattacharyya}}, \bibinfo {author} {\bibfnamefont {M.}~\bibnamefont {Banerjee}}, \bibinfo {author} {\bibfnamefont {H.}~\bibnamefont {Nhalil}}, \bibinfo {author} {\bibfnamefont {S.}~\bibnamefont {Islam}}, \bibinfo {author} {\bibfnamefont {C.}~\bibnamefont {Dasgupta}}, \bibinfo {author} {\bibfnamefont {S.}~\bibnamefont {Elizabeth}},\ and\ \bibinfo {author} {\bibfnamefont {A.}~\bibnamefont {Ghosh}},\ }\href {https://doi.org/10.1021/acsnano.5b06163} {\bibfield  {journal} {\bibinfo  {journal} {ACS Nano}\ }\textbf {\bibinfo {volume} {9}},\ \bibinfo {pages} {12529} (\bibinfo {year} {2015})},\ \bibinfo {note} {pMID: 26549529},\ \Eprint {https://arxiv.org/abs/https://doi.org/10.1021/acsnano.5b06163} {https://doi.org/10.1021/acsnano.5b06163} \BibitemShut {NoStop}%
\bibitem [{\citenamefont {Horák}, \citenamefont {Čermák},\ and\ \citenamefont {Koudelka}(1986)}]{linear_antisite}%
  \BibitemOpen
  \bibfield  {author} {\bibinfo {author} {\bibfnamefont {J.}~\bibnamefont {Horák}}, \bibinfo {author} {\bibfnamefont {K.}~\bibnamefont {Čermák}},\ and\ \bibinfo {author} {\bibfnamefont {L.}~\bibnamefont {Koudelka}},\ }\href {https://doi.org/https://doi.org/10.1016/0022-3697(86)90010-7} {\bibfield  {journal} {\bibinfo  {journal} {Journal of Physics and Chemistry of Solids}\ }\textbf {\bibinfo {volume} {47}},\ \bibinfo {pages} {805} (\bibinfo {year} {1986})}\BibitemShut {NoStop}%
\bibitem [{\citenamefont {Lo{\v{s}}t'{\'a}k}, \citenamefont {Karamazov},\ and\ \citenamefont {Horak}(1994)}]{antisite_vacancies}%
  \BibitemOpen
  \bibfield  {author} {\bibinfo {author} {\bibfnamefont {P.}~\bibnamefont {Lo{\v{s}}t'{\'a}k}}, \bibinfo {author} {\bibfnamefont {S.}~\bibnamefont {Karamazov}},\ and\ \bibinfo {author} {\bibfnamefont {J.}~\bibnamefont {Horak}},\ }\href@noop {} {\bibfield  {journal} {\bibinfo  {journal} {physica status solidi (a)}\ }\textbf {\bibinfo {volume} {143}},\ \bibinfo {pages} {271} (\bibinfo {year} {1994})}\BibitemShut {NoStop}%
\bibitem [{\citenamefont {Mihaila}, \citenamefont {Dinulescu},\ and\ \citenamefont {Varasteanu}(2023)}]{phonons}%
  \BibitemOpen
  \bibfield  {author} {\bibinfo {author} {\bibfnamefont {M.}~\bibnamefont {Mihaila}}, \bibinfo {author} {\bibfnamefont {S.}~\bibnamefont {Dinulescu}},\ and\ \bibinfo {author} {\bibfnamefont {P.}~\bibnamefont {Varasteanu}},\ }\href {https://doi.org/10.1063/5.0144474} {\bibfield  {journal} {\bibinfo  {journal} {Applied Physics Letters}\ }\textbf {\bibinfo {volume} {122}},\ \bibinfo {pages} {174003} (\bibinfo {year} {2023})}\BibitemShut {NoStop}%
\bibitem [{\citenamefont {Serrano-S{\'a}nchez}\ \emph {et~al.}(2017)\citenamefont {Serrano-S{\'a}nchez}, \citenamefont {Gharsallah}, \citenamefont {Nemes}, \citenamefont {Biskup}, \citenamefont {Varela}, \citenamefont {Mart{\'\i}nez}, \citenamefont {Fern{\'a}ndez-D{\'\i}az},\ and\ \citenamefont {Alonso}}]{phonon2}%
  \BibitemOpen
  \bibfield  {author} {\bibinfo {author} {\bibfnamefont {F.}~\bibnamefont {Serrano-S{\'a}nchez}}, \bibinfo {author} {\bibfnamefont {M.}~\bibnamefont {Gharsallah}}, \bibinfo {author} {\bibfnamefont {N.}~\bibnamefont {Nemes}}, \bibinfo {author} {\bibfnamefont {N.}~\bibnamefont {Biskup}}, \bibinfo {author} {\bibfnamefont {M.}~\bibnamefont {Varela}}, \bibinfo {author} {\bibfnamefont {J.}~\bibnamefont {Mart{\'\i}nez}}, \bibinfo {author} {\bibfnamefont {M.}~\bibnamefont {Fern{\'a}ndez-D{\'\i}az}},\ and\ \bibinfo {author} {\bibfnamefont {J.}~\bibnamefont {Alonso}},\ }\href@noop {} {\bibfield  {journal} {\bibinfo  {journal} {Scientific reports}\ }\textbf {\bibinfo {volume} {7}},\ \bibinfo {pages} {6277} (\bibinfo {year} {2017})}\BibitemShut {NoStop}%
\bibitem [{\citenamefont {Shklovskii}\ and\ \citenamefont {Efros}(2013)}]{shklovskii2013electronic}%
  \BibitemOpen
  \bibfield  {author} {\bibinfo {author} {\bibfnamefont {B.~I.}\ \bibnamefont {Shklovskii}}\ and\ \bibinfo {author} {\bibfnamefont {A.~L.}\ \bibnamefont {Efros}},\ }\href@noop {} {\emph {\bibinfo {title} {Electronic properties of doped semiconductors}}},\ Vol.~\bibinfo {volume} {45}\ (\bibinfo  {publisher} {Springer Science \& Business Media},\ \bibinfo {year} {2013})\BibitemShut {NoStop}%
\bibitem [{\citenamefont {Sharma}\ \emph {et~al.}(2025)\citenamefont {Sharma}, \citenamefont {R}, \citenamefont {Gopal},\ and\ \citenamefont {Mitra}}]{sharma2025bandmeanderingchargedimpurity}%
  \BibitemOpen
  \bibfield  {author} {\bibinfo {author} {\bibfnamefont {K.}~\bibnamefont {Sharma}}, \bibinfo {author} {\bibfnamefont {N.~K.}\ \bibnamefont {R}}, \bibinfo {author} {\bibfnamefont {R.~K.}\ \bibnamefont {Gopal}},\ and\ \bibinfo {author} {\bibfnamefont {C.}~\bibnamefont {Mitra}},\ }\href {https://arxiv.org/abs/2507.08579} {\enquote {\bibinfo {title} {Band meandering due to charged impurity effects and carrier transport in ternary topological insulators},}\ } (\bibinfo {year} {2025}),\ \Eprint {https://arxiv.org/abs/2507.08579} {arXiv:2507.08579 [cond-mat.mtrl-sci]} \BibitemShut {NoStop}%
\end{thebibliography}%

\end{document}